\documentclass[bibyear]{aa}  

\usepackage{xcolor}
\usepackage{graphicx}
\usepackage{natbib}
\usepackage{txfonts}
\usepackage[hidelinks]{hyperref}
\hypersetup{colorlinks=true,linkcolor=blue,citecolor=blue,urlcolor=blue}
\usepackage{placeins}
\usepackage{etoolbox}
\AtBeginDocument{\let\linenumbers\nolinenumbers}

\begin{document}

   \title{Surveying exogenous species in Saturn with ALMA}

   \subtitle{I. Detecting and Mapping CO}

   \author{Deborah Bardet \inst{1} \corrauth{deborah.bardet@obspm.fr}
            \and
        Thierry Fouchet \inst{1} \email{thierry.fouchet@obspm.fr}
            \and
        Thibault Cavali\'{e}\inst{2,1} \email{thibault.cavalie@u-bordeaux.fr}
            \and
        Raphaël Moreno\inst{1} \email{raphael.moreno@obspm.fr}
            \and
        Emmanuel Lellouch\inst{1} \email{emmanuel.lellouch@obspm.fr}
            \and
        Camille Lefour\inst{2} \email{camille.lefour@u-bordeaux.fr}
            \and
        Bilal Benmahi\inst{3,4} \email{billal.benmahi@lam.fr}
            \and
        Sandrine Guerlet\inst{5,1} \email{sandrine.guerlet@lmd.ipsl.fr}
          }

   \institute{LIRA, Observatoire de Paris, Université PSL, CNRS, Sorbonne Université, 
                    Université Paris Cité, CY Cergy Paris Université, 5 place Jules
                    Janssen, 92195 Meudon, France
         \and
             Laboratoire d'Astrophysique de Bordeaux, Univ. Bordeaux, 
                    CNRS, B18N, allée Geoffroy Saint-Hilaire, 33615 Pessac, France
        \and
            Aix-Marseille Université, CNRS, CNES, Institut Origines, LAM, Marseille, France
        \and
            Laboratory of Atmospheric and Planetary Physics, STAR Institute, University of Liège, Belgium
        \and
            Laboratoire de M\'{e}t\'{e}orologie Dynamique / Institut Pierre-Simon Laplace (LMD/IPSL), CNRS, Sorbonne Universit\'{e}, \'{E}cole Polytechnique, Institut Polytechnique de Paris, \'{E}cole Normale Sup\'{e}rieure (ENS), PSL Research University, Campus Pierre et Marie Curie BC99, 4 place Jussieu, 75005 Paris, France
        }

   \date{}

% \abstract{}{}{}{}{} 
% 5 {} token are mandatory
 
  \abstract
   {The origin of carbon monoxide (CO) in Saturn’s stratosphere remains uncertain, with proposed sources including internal thermochemical production, cometary impacts, and exogenic material from the rings and icy moons (i.e. Enceladus).
   }
   {We aim to constrain the vertical and meridional distribution of stratospheric CO and assess the relative contributions of these potential sources.}
   {Here, we analysed high-spectral-resolution ALMA observations of the CO (J=3–2) line obtained on 25 May 2018, sampling Saturn’s limb from 20$^\circ$S to 70$^\circ$N. We retrieved CO vertical profiles using a line-by-line radiative transfer model combined with spectral inversion techniques, testing multiple prior scenarios representative of different source hypotheses.}
   {Stratospheric CO is confined to a narrow layer between 0.1 and 10 mbar, with a robust negative vertical gradient from a mean abundance of $(2.0\pm0.7)\times10^{-8}$ (in volume mixing ratio) at 0.1 mbar to a mean value of $(4.3\pm1.5)\times10^{-8}$ at 1 mbar, followed by a robust positive vertical gradient up to a mean value of $(5.1\pm1.7)\times10^{-9}$ at 10 mbar. The meridional distribution is statistically homogeneous, with a marginal enhancement near 60$^\circ$N plausibly related to Enceladus. No significant equatorial enhancement is detected.}
   {The absence of a strong equatorial enhancement rules out a long-lived steady source associated with ring influx. The observations are most consistent with a combination of exogenous materials: a cometary impact, which causes the vertical diffusion of CO, and the injection of additional water from the rings and/or Enceladus near 60$^\circ$N, which, through photolysis, converts CO into CO$_2$ at high altitude.}

   \keywords{Planets and satellites: individual: Saturn – Planets and satellites: atmospheres – Techniques: imaging spectroscopy
               }

   \maketitle
%
%-------------------------------------------------------------------
\section{Introduction}
Saturn’s atmosphere contains exogenous materials such as water, hydrogen cyanide, carbon dioxide, and carbon monoxide \citep{Feuc:97,Fouc:26,Cava:09}. 
Assessing the origin of carbon monoxide in Saturn's middle atmosphere remains, however, an open question. 
In addition to an internal source due to thermochemical processes \citep{Noll:91,Cava:24}, multiple external sources might deliver material at the top of the stratosphere, and separating their individual contributions is a challenge for spectroscopic analysis. 
On Jupiter \citep{Beza:02} and Neptune \citep{Lell:05,Hesm:07}, stratospheric CO primarily originates from external sources, namely large cometary impacts \citep[the most recent one being the collision of comet Shoemaker-Levy 9, SL9, with Jupiter in 1994,][]{Lell:95,Lell:97,More:03,Grif:04}. 
In the same vein, previous submillimetric observations of CO in the stratosphere of Saturn were also found to be consistent with a cometary source that would have impacted the planet two centuries ago \citep{Cava:08giant,Cava:09,Cava:10}. 
Photochemical modelling of Saturn's stratosphere later supported this scenario: \cite{Mose:17} showed the necessity of a large amount of external oxygen from large cometary impacts within the last few hundred years to explain the observed CO abundance at Saturn.

Nevertheless, Saturn's rings and icy moons constitute additional potential sources of exogenic material at the top of the  stratosphere, mostly in the form of water but also in the form of carbon-bearing molecules, and hence might also alter the global distribution of Saturn's CO \citep[e.g., from water photochemistry,][]{Mose:00PartI,Mose:00PartII}. 
Indeed, in Saturn's stratosphere, externally provided oxygen-bearing and carbon-bearing molecules may be photochemically converted into CO by different pathways, such as the indirect pathways initiated by H$_2$O photolysis, followed by the reaction of O and OH with CH$_3$ \citep{Mose:23ringrain}. 
In this context, monitoring oxygen-bearing molecules such as water is as important as monitoring CO itself.
For instance, \cite{Cava:19SaturnH2O} demonstrated that Saturn's stratospheric water is not distributed uniformly with latitude, showing a maximum abundance at the equator.
This suggests the primary source of Saturn’s stratospheric H$_2$O lies in the equatorial plane, and \cite{Cava:19SaturnH2O} identified Enceladus plumes as dominating the flux. 
Cassini's Grand Finale orbits ---the closest approach to Saturn between the inner D ring and the equatorial atmosphere--- allowed in-situ measurements of material infalling from the rings \citep[such as CH$_4$, NH$_3$, H$_2$O, CO, N$_2$, CO$_2$, silicates, etc., ][]{Perr:18,Hsu:18,Mitc:18,Wait:18,Mill:20SaturnRingFalling,Seri:22SaturnRings}, onto the equatorial upper atmosphere and interacting with it. 
In addition to providing water in the equatorial plane, Enceladus and the rings could provide materials to Saturn's mid-latitudes. 
Indeed, once ionized, water released in Saturn's magnetosphere plunges into the ionosphere, following the magnetic field lines at specific latitudes magnetically conjugated with the rings and Enceladus \citep[a mechanism named ``ring rain'' for material exclusively coming from the rings,][]{Conn:86,ODon:19SaturnRingRain}. 
Throughout this so-called ``ring rain'' process, the Saturn rings provide additional exogenic material for the production of CO as well \citep{Moor:15SaturnRingRain,Wait:18,ODon:19SaturnRingRain}.  
Observations of the variation in the density of H$_3^+$ as a function of latitude suggested that, together with water from Enceladus plumes, the ``ring rain'' exogenic material primarily influences the local ionospheric chemistry at specific latitudes and might produce carbon monoxide during photochemical reactions.

Estimating the individual contributions of CO and H$_2$O originating from Enceladus, the rings, and cometary impacts is challenging, and a wide range of values has been reported to date. 
\cite{Cava:10} estimated that a cometary impact about two centuries ago might have brought about 2.5$\times$10$^{12}$ kg of CO into the stratosphere of Saturn. 
Considering the ``ring rain'' mechanism, several global ionized water flux estimations exist to date, ranging from 3-20 kg/s \citep[][(1.0-6.8)$\times$10$^{26}$ molecule s$^{-1}$ in the original text]{Moor:15SaturnRingRain} to 432-2870 kg/s \citep{ODon:19SaturnRingRain}.
Cassini Grand Finale observations of the rings influx into the equatorial stratosphere showed N$_2$ and CO dominance \citep{Seri:22SaturnRings}, as well as H$_2$O, CH$_4$, NH$_3$, and CO$_2$ \citep{Mill:20SaturnRingFalling}.  
While \cite{Mitc:18}, \cite{Wait:18} and \cite{Mill:20SaturnRingFalling} estimated a total influx of material of 5 kg/s (for atomic components), 4,800-45,000 kg/s and (1.7-8.8)$\times$10$^{5}$ kg/s, respectively, \cite{Seri:22SaturnRings} were able to differentiate the flows of the main atomic components and estimated a CO influx of 7000 kg/s, a H$_2$O influx of 4400 kg/s, and a CO$_2$ influx of 2000 kg/s, all falling into the equatorial stratosphere. 
All the estimates listed above cover a wide range of orders of magnitude: from very small quantities to quantities so large that they seem unlikely. 
It is therefore crucial to ascertain what the actual situation is.

Each of the aforementioned external sources occurs in a specific latitudinal range. 
For instance, observations of the meridional density of H$_3^+$ \citep{Moor:15SaturnRingRain,ODon:19SaturnRingRain} depicted local density enhancements at $\pm$40$^\circ$ and $\pm$50$^\circ$ resulting from material inputs from the ``ring rain'', consistent with observations of hydrocarbons in the ultraviolet spectrum \citep{Pran:06}. 
Enceladus plumes provide water at Saturn's mid-latitudes: at 62$^\circ$S \citep{Moor:15SaturnRingRain} and at 65$^\circ$N \citep{Conn:86} in the form of charged particles following the magnetic field lines; as well as at the equator by direct influx in neutral form in the equatorial plane \citep[][direct influx which is added and mixed to the infalling material from the rings measured during the Grand Finale]{Cava:19SaturnH2O}. 
Given the distinct latitudinal patterns associated with the different sources, one might expect that the meridional variations of CO could provide clues about the dominating source. 
This statement  must, however, be tempered by our lack of observational constraints on Saturn's stratospheric meridional circulation \citep[solely model predictions have been provided by][]{Bard:22} and by how efficiently this circulation transports material relative to the typical CO photochemical production and destruction timescales.

Previous observations of Saturn's stratospheric CO \citep{Cava:09,Cava:10} did not cover the entire latitude range and lacked the spectral and angular resolutions required to assess the vertical and meridional structures of the CO distribution simultaneously. 
In this paper, we present stratospheric CO abundances retrieved from a submillimetric dataset of CO observations.
This dataset has been obtained using the Atacama Large Millimeter/submillimeter Array facility, potentially shedding light on the contribution of CO external or internal sources to its stratospheric vertical and meridional distribution. 

Hereafter, we present the observations in section \ref{sec:obs}. 
We describe the spectral inversion model used to retrieve the CO profiles at each latitude, as well as the prior CO profiles tested, the reference atmosphere prescribed to the model, and the determination of vertical sensitivity in section \ref{sec:model}. 
We discuss the retrieved CO meridional and vertical distributions in section \ref{sec:results} and evaluate the potential origin(s) of CO in section \ref{sec:discussion_equatorial_origin}.
Finally, we conclude in section \ref{sec:ccl}.

\section{Observations}
\label{sec:obs}
Using the Atacama Large Millimeter/submillimeter Array (ALMA, project 2017.1.00636.S) on May 25$^\mathrm{th}$ 2018, we have obtained maps of HCN (J=5-4) and CO (J=3-2) lines with a combination of seven pointings of the main array in configuration C43.2, complemented by three pointings of the Atacama Compact Array (ACA) to mosaic the full disc of the planet.
This dataset has been designed using 43 operating antennas, resulting in a spatial resolution of the main array of 0.63''$\times$0.51''.
Given the equatorial apparent size of Saturn around 18.04'' at the date of observation, the resulting latitudinal resolution is $\approx$5$^\circ$ from the equatorial region to mid-latitudes, and $\approx$10$^\circ$ at polar latitudes.
The sub-Earth latitude was 30.54$^\circ$N at the date of observation, leading to a limb scanning from 69$^\circ$N to 50$^\circ$S (the north pole is visible on the data cube, but it is not located on the limb of the planet due to the sub-Earth latitude). 
For both HCN (J=5-4) and CO (J=3-2), the dataset is composed of two spectral windows centred on the line rest frequency: a narrow spectral window of 0.5 GHz (for HCN) and 0.120 GHz (for CO) width, and a large bandwidth of 2.0 GHz width.  
Both narrow and large spectral windows are sampled with 2048 channels, implying a spectral resolution of 70 kHz for the narrow window, and a spectral resolution of 1.1 MHz for the large one.

\begin{figure}
    \centering
    \includegraphics[width=\linewidth]{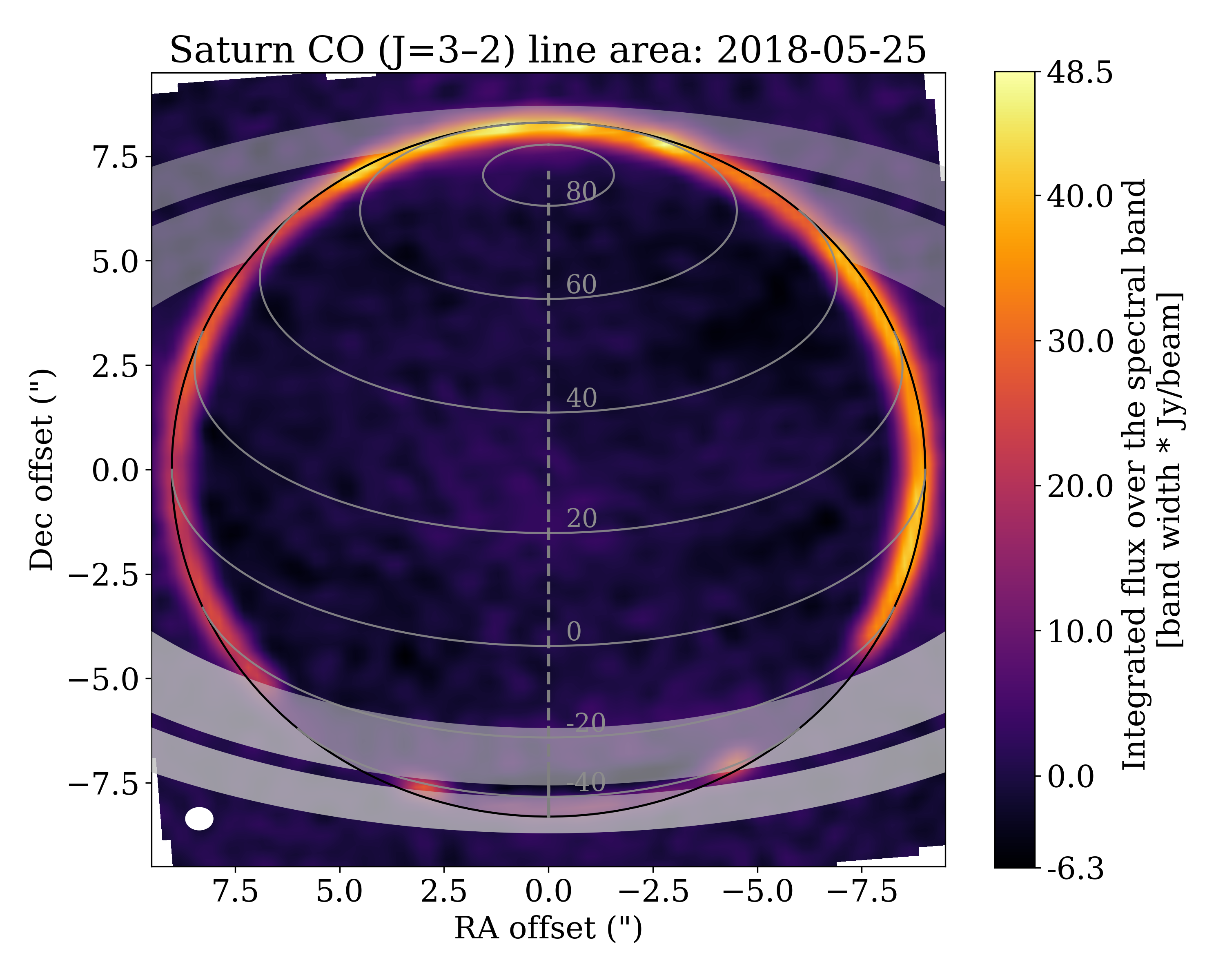}
    \caption{Map of the CO (J=3-2) line area, as observed with ALMA on May 25$^\mathrm{th}$, 2018, after reduction, correction, calibration and subtraction of the continuum image (see \citealt{Benm:22} for method details). 
    The black ellipse represents the the planetary limb at the 1 bar level. 
    The central meridian is shown with a grey dashed line. Latitude parallels are displayed with grey solid lines. The position of the A and B rings are highlighted with a grey-filled region. The beam is illustrated with a white filled ellipse in the bottom left corner of the map.}
    \label{fig:CO_line_intensity_map}
\end{figure}

\begin{figure}
    \centering
    \includegraphics[width=\linewidth]{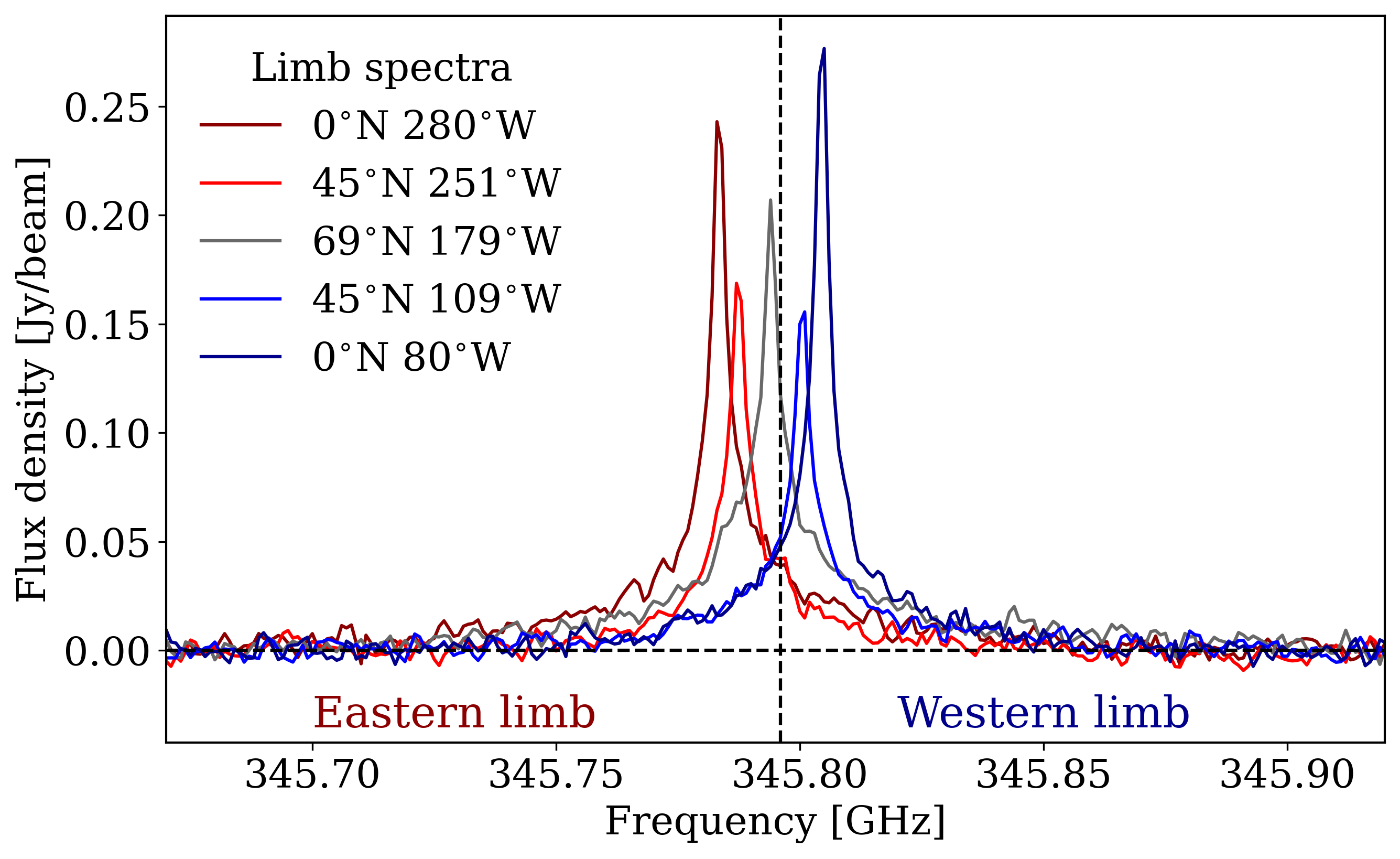}
    \caption{Broadband limb continuum-subtracted CO spectra extracted from the image presented in Fig. \ref{fig:CO_line_intensity_map} at eastern and western limbs for different latitudes (red to blue lines). 
    The CO (J=3-2) rest frequency is indicated by the vertical black dashed line (345.796 GHz) and the zero flux density is represented by the horizontal black dashed line. From the northernmost latitude of 70$^{\circ}$N to the equator, the intensity and the width of the limb CO line varies depending on latitude, implying variation on the vertical information content. 
    }
    \label{fig:raw_data_line_comparison}
\end{figure}

Together with the HCN line, the CO line observed by ALMA was detected at the limb of the planet, from the equatorial region to the northern polar regions (Figure \ref{fig:CO_line_intensity_map}). 
Those two datasets were initially used to provide the first absolute wind measurements in Saturn's stratosphere \citep{Benm:22,Benm:25}, by retrieving the Doppler shift of the observed HCN and CO lines from their rest frequency.
A detailed description of the data reduction, calibration, and processing is therefore available in \cite{Benm:22}.
From the emission map of the CO (J=3-2) line centred at 345.796 GHz (Figure \ref{fig:CO_line_intensity_map}), these authors produced a CO dataset composed of 427 spectra along the limb, as the sampling between two successive spectrum positions was one-fourth of the size of the ALMA synthetic beam ($\approx$0.13'').
Because the sampling interval is smaller than the beam, each spectrum is not independent from its closest neighbour. 
Therefore, in our retrieval, we will consider spatial features only if their size is larger than the initial resolution of the ALMA main array (i.e., $\approx$5$^\circ$ from the equatorial region to mid-latitudes, and $\approx$10$^\circ$ at polar latitudes).  
In addition to retrieving the zonal wind velocity, those previous analyses revealed variations in the intensity and width of the CO line across the entire limb, suggesting that information on the vertical and horizontal distribution of Saturn’s CO is accessible (see Figure \ref{fig:raw_data_line_comparison}): this is the aim of the present study.
We also draw the reader's attention to a companion paper in which we present a parallel study dedicated to the retrieval of the HCN spatial distribution \citep{Fouc:26}.

In May 2018, Saturn's atmosphere was under northern summer conditions, and its ring system obscured the southern hemisphere (since it is optically thick). 
Therefore, there is not enough observational information for latitudes south of 20$^{\circ}$S.
In the following result sections (sections \ref{sec:results} and \ref{sec:discussion_equatorial_origin}), we therefore present retrieval results of CO abundances only from 20$^{\circ}$S to 70$^{\circ}$N. 
All analysed spectra in this paper are obtained from the continuum-subtracted data cubes (Figure \ref{fig:CO_line_intensity_map}).

\begin{figure}
    \includegraphics[width=\linewidth]{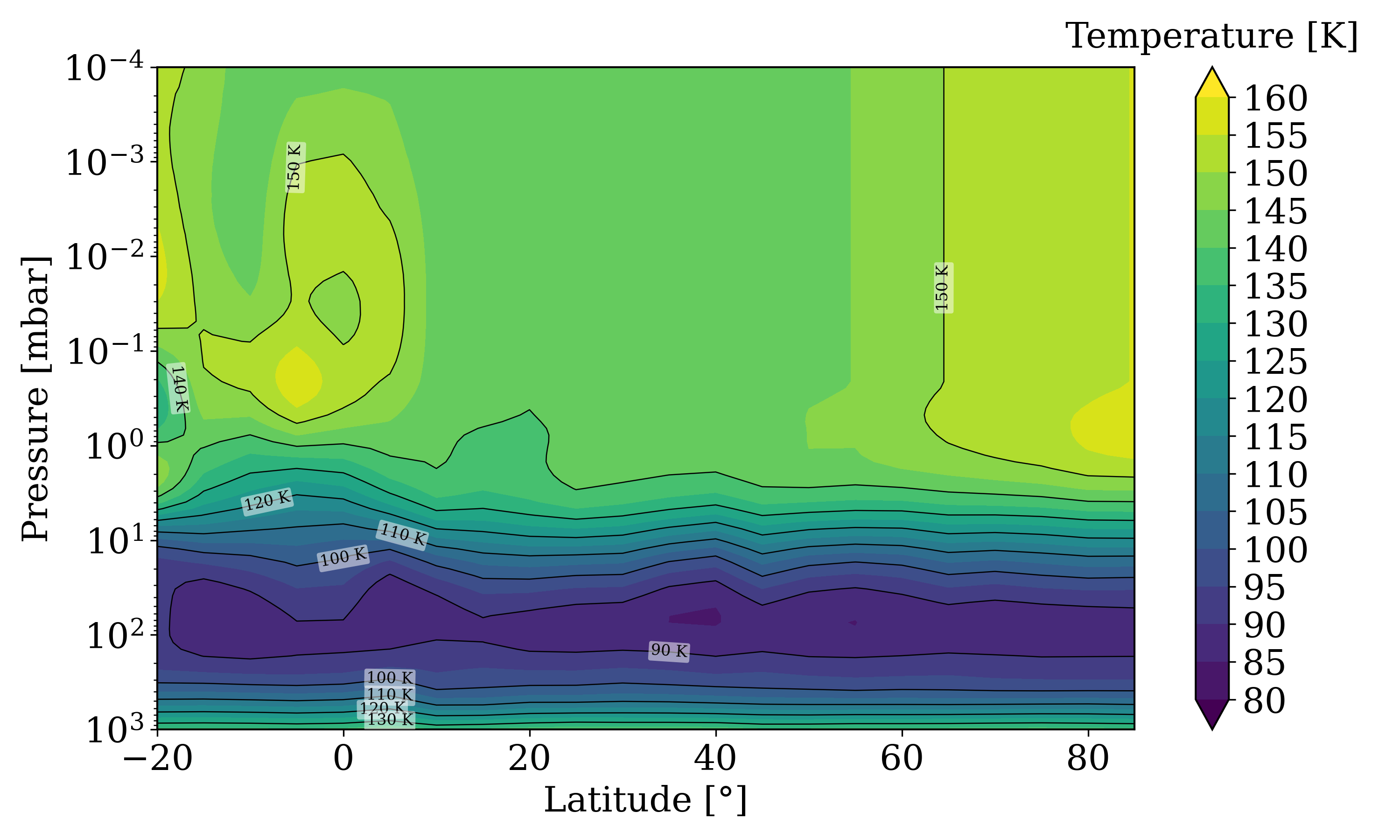}
    \caption{Temperature field used as input by the line-by-line radiative transfer model. This is a temperature map created from a combination of Cassini/CIRS observations in limb view from 15$^\circ$S to 10$^\circ$N, and in nadir view out of this equatorial region.}
    \label{fig:temperature_inputs}
\end{figure}

\section{Spectral modelling}
\label{sec:model}
\subsection{Radiative transfer model}
We forward-modelled the spectroscopic observations with the line-by-line radiative transfer model described in \cite{Cava:19SaturnH2O}, and combined it with spectral inversion calculations \citep{Guer:09,Fouc:16} to obtain vertical profiles of the CO abundance along the limb. 
The forward model accounts for the 3D ellipsoidal geometry of the planet and includes the emission and absorption of the main rings, with variable ring brightness temperature and absorption coefficients \citep[see details in ][]{Cava:19SaturnH2O}. 
Collision-induced absorption from H$_2$-H$_2$, H$_2$-He, and H$_2$-CH$_4$ is also included in the radiative transfer calculations following \cite{Bory:85}, \cite{Bory:86}, and \cite{Bory:88}.  
In addition, the relevant H$_2$/He broadening parameters have been calculated for PH$_3$ \citep{Levy:93, Levy:94}, for NH$_3$ \citep{Flet:07SaturnPH3}, and for CO \citep{Dick:09_CO}).

\subsection{Thermal field}
Accurate knowledge of the thermal field, with similar meridional coverage as the ALMA CO line observations and sufficient vertical resolution, is crucial to the retrieval of CO profiles. 
We used a combination of temperature profiles retrieved from the Cassini Composite Infrared Spectrometer (CIRS) in 2016 and 2017 to create a pressure-latitude temperature map that we assumed to represents Saturn's temperature structure in 2018 well enough, given its long radiative timescales (Figure \ref{fig:temperature_inputs}).
We used the temperature retrieved from observations in limb viewing from 15$^\circ$S to 10$^\circ$N \citep[observation dates from  November 2016 to August 2017,][]{Brow:24}, combined with temperature retrieved from observations in nadir view outside of this specific equatorial region, originally processed by \cite{Flet:18hexagon} (nadir observation date is May 20$^\mathrm{th}$ 2017).
The CIRS instrument sensed a wavelength range of 600--1400 cm$^{-1}$, probing in nadir geometry the 0.2--5 mbar and 70--250 mbar pressure ranges.
For pressures lower than 0.2 mbar, we chose to extrapolate the temperature profiles as isothermally, since the temperature profiles retrieved by \cite{Flet:18hexagon} return to the prior temperature profiles whose validity is questionable.  
However, the limb view geometry in the equatorial region enables both a higher vertical resolution and a broader vertical extent, which allows a better characterisation of Saturn's equatorial oscillation \citep{Orto:08,Fouc:08,Guer:11} as the vertical sensitivity in the temperature retrieval is extended to 10$^{-2}$ mbar.

\subsection{Reference atmosphere composition and CO prior profiles}

The reference atmosphere composition used for our radiative transfer calculations is similar to that used by \cite{Cava:19SaturnH2O} and \cite{Lefo:25} for H$_2$, He, CH$_4$, NH$_3$ and PH$_3$ vertical distribution. 
We therefore used the vertical distributions of NH$_3$ and PH$_3$ from \cite{Davi:96}
and \cite{Flet:09JupSatPH3}, respectively.

\begin{figure}
    \centering
    \includegraphics[width=0.7\linewidth]{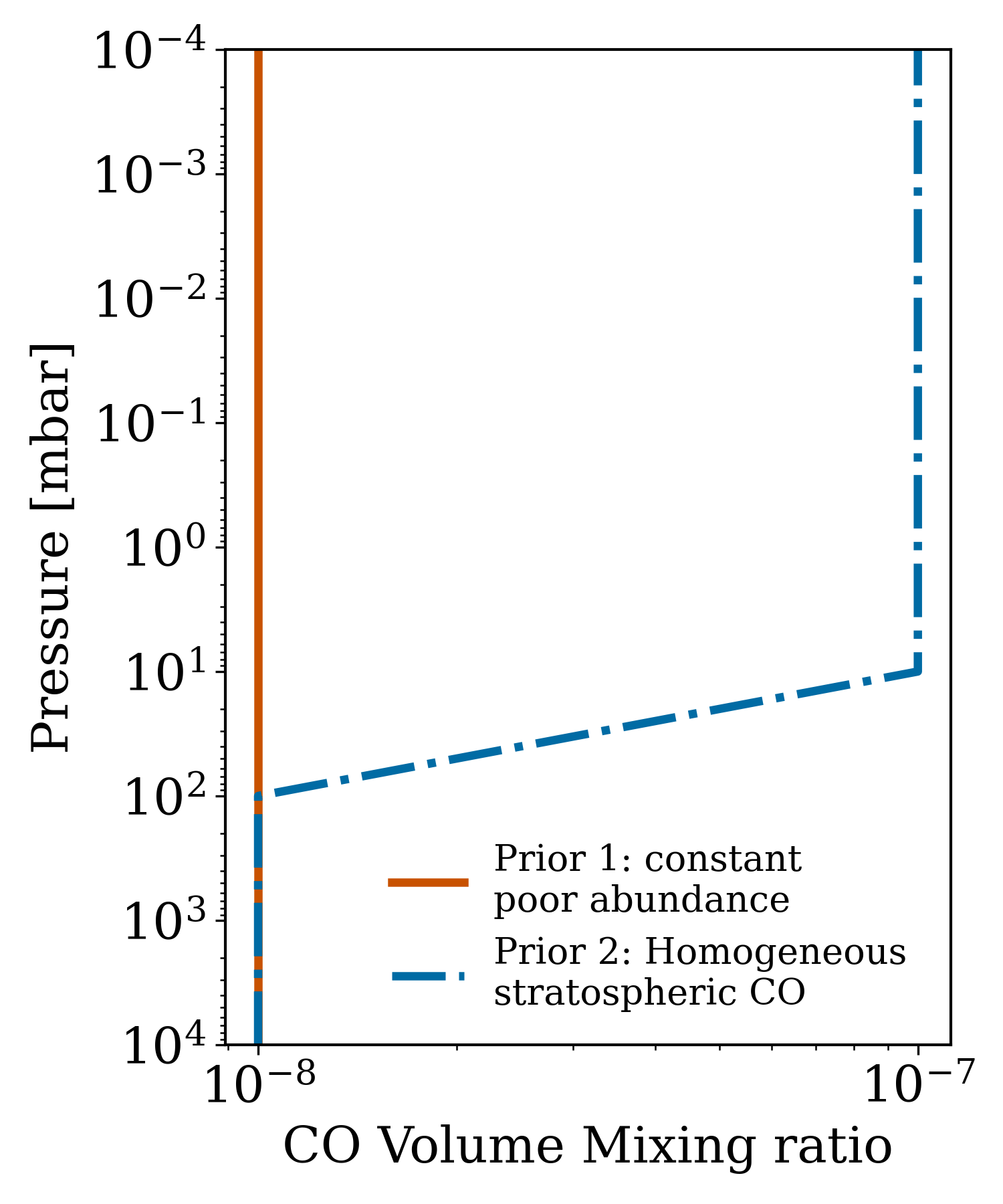}
    \caption{CO prior profiles used in this study: Prior 1 (dark orange) is a constant low abundance profile set at a volume mixing ratio of 10$^{-8}$, Prior 2 (blue, dash-dotted line) is a two-layered profile describing an homogeneous stratospheric CO originated from an external source scenario.}
    \label{fig:CO_priors}
\end{figure}

To explore a variety of possible CO sources in Saturn's stratosphere, we tested two different prior CO abundance profiles in our retrieval calculations.
Displayed in Figure \ref{fig:CO_priors}, Prior 1 is a simple profile describing a vertically-homogeneous CO abundance over the entire pressure range simulated, with a low abundance set at a volume mixing ratio of 10$^{-8}$. 
In addition, we explored the external source scenario. 
To that end, we added one CO prior profile (Prior 2) describing a two-layer vertical structure. 
Prior 2 represents the case of a predominantly external source, with a stratospheric mixing ratio (above the 10 mbar pressure level) 10 times larger than the tropospheric one.

\subsection{Inversion model}
Our forward-modelled atmosphere ranges from 10$^{4}$ to 10$^{-4}$ mbar, discretized on a grid of 1000 vertical levels. 
The CO vertical distribution was retrieved using a constrained, regularized inversion scheme based on the approach originally developed by \cite{Conr:00Saturn} and subsequently described by \cite{Fouc:16} and \cite{Lell:17Pluton}. 
The method relies on a linearization of the radiative transfer with respect to the CO profile, yielding the Jacobian matrix $K$, and on the minimization of a cost function that simultaneously accounts for the agreement with the observed spectra, the deviation from the prior profile, and the vertical smoothness of the solution. 
This regularization effectively suppresses spurious oscillations at the pressure levels constrained by the observations, and prevents unrealistic values as much as possible where the measurements provide little or no sensitivity. All retrievals were performed assuming a vertical correlation length of 1.5 scale heights, over a grid of 33 vertical levels sampled between 10$^{4}$ to 10$^{-4}$ mbar.

\subsection{Vertical Sensitivity}
\label{SecVerticalResults}

Preliminary retrieval calculations using the narrow-bandwidth dataset of the CO (J=3-2) line (with a resolution of 70 kHz) failed to properly constrain the vertical profile (see Figure \ref{fig:app_compare_high-low_resolution_profiles}). 
Together with the Doppler shift of the line from one limb to the other, the large width of the CO line compared to the observational bandwidth prevented us from obtaining a zero-flux density while calculating continuum-subtracted CO spectra at the edge of the narrow frequency bandwidth (Figure \ref{fig:app_compare_high-low_resolution_spectrum}).
Hence, at each inversion iteration, the synthetic spectrum was always slightly offset from this zero-flux density level, degrading the accuracy of the Jacobian derivatives in the inversion calculation and, thus, the sensitivity pressure range of the observations. 
In what follows, we make use of the large bandwidth dataset, scanning a frequency range from 343.978 to 345.977 GHz, with a frequency resolution of 1.1 MHz. 
Despite the choice of a wide bandwidth, we do not lose any significant spectral signal (see Figures \ref{fig:app_contrib_narrowband} and \ref{fig:app_contrib_broadband}).

\begin{figure}
    \centering
    \includegraphics[width=\linewidth]{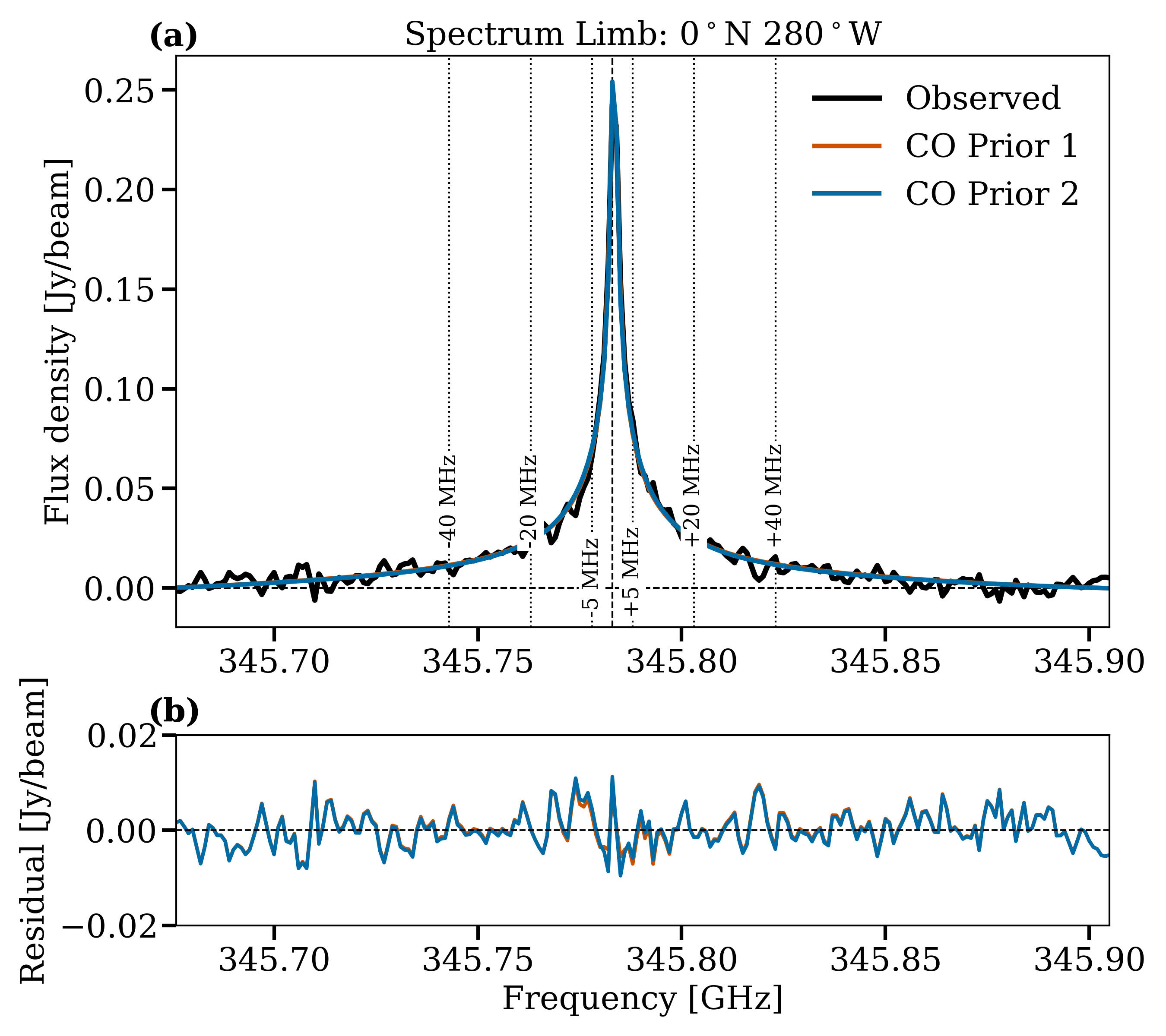}
    \caption{Example of the CO line fitting  for (a) the eastern equatorial limb spectrum (at 0$^{\circ}$N-280$^\circ$W) for the two retrieval calculations carried in this study, and (b) the residual flux density (observed spectrum - synthetic spectrum) showing that retrievals starting from the 2 priors result in fitting the observed limb spectrum within the noise of the observations. The black dashed line indicates the frequency of the peak, and the dotted lines depict different values of frequency offset from the peak, for which we have calculated the contribution function.}
    \label{fig:comparison_spectre423}
\end{figure}

\begin{figure*}
    \centering
    \includegraphics[width=\linewidth]{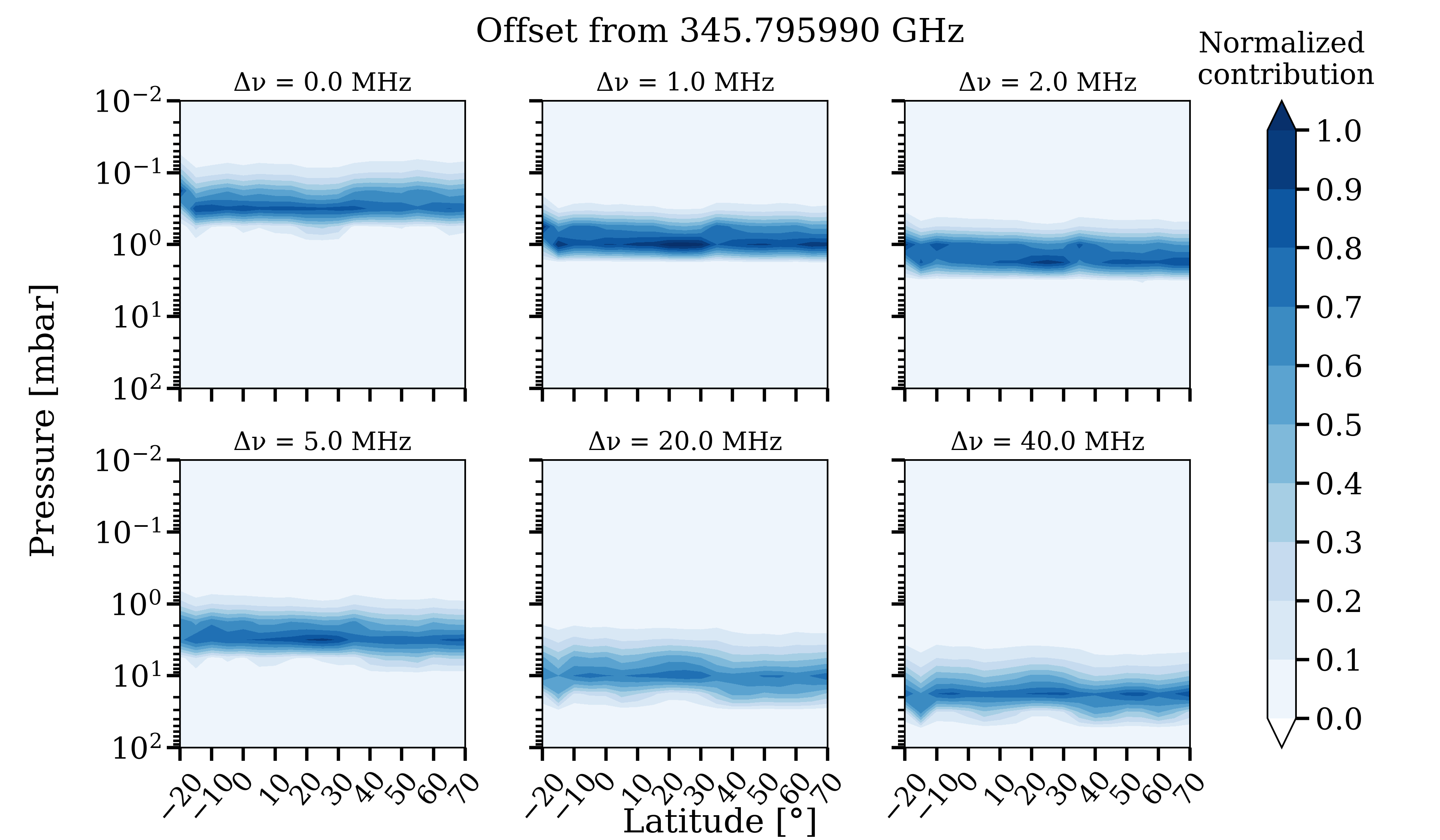}
    \caption{Normalized contribution function for Prior 1, calculated for six frequencies at and around the CO (J=3-2) line core frequency. Those frequencies are materialised with dashed and dotted lines on Figure \ref{fig:comparison_spectre423}. As the forward-model of \cite{Cava:19SaturnH2O} accounts for the Doppler shift induced by the planet rotation, those contribution function are valid all along the limb. 
    }
    \label{fig:contribution_function}
\end{figure*}
Figure \ref{fig:comparison_spectre423} presents an example of the CO line fitting at the eastern equator starting from the two CO prior profiles tested in this study, along with the corresponding residual flux densities. 
Both inversions successfully reproduce the observed spectrum to within the noise level of the data. 
Additional examples can be found in Appendix \ref{app:line_fitting} for different latitude points. 
Meridional variations of the contribution functions for CO retrieved from Prior 1 are presented for six frequencies within the core or the wings of the CO (J=3-2) line in Figure \ref{fig:contribution_function}, depicting latitudinal and pressure trend of the maximum vertical sensitivity.    
This trend is consistent across the two CO priors considered to calculate the contribution function (see normalized contribution functions for CO prior 2 in Figure \ref{fig:app_contribution_function_prior2}). 
At the line core frequency (345.796 GHz), the data are mainly sensitive to the 0.2 mbar pressure level. 
However, the maximum contribution function lies between $\pm$ 1 and 2 MHz from the maximum of the CO line, with the sensitivity of the data to pressures ranging from 0.7 to 1 mbar.

\begin{figure*}
   \centering
   \includegraphics[width=0.9\linewidth]{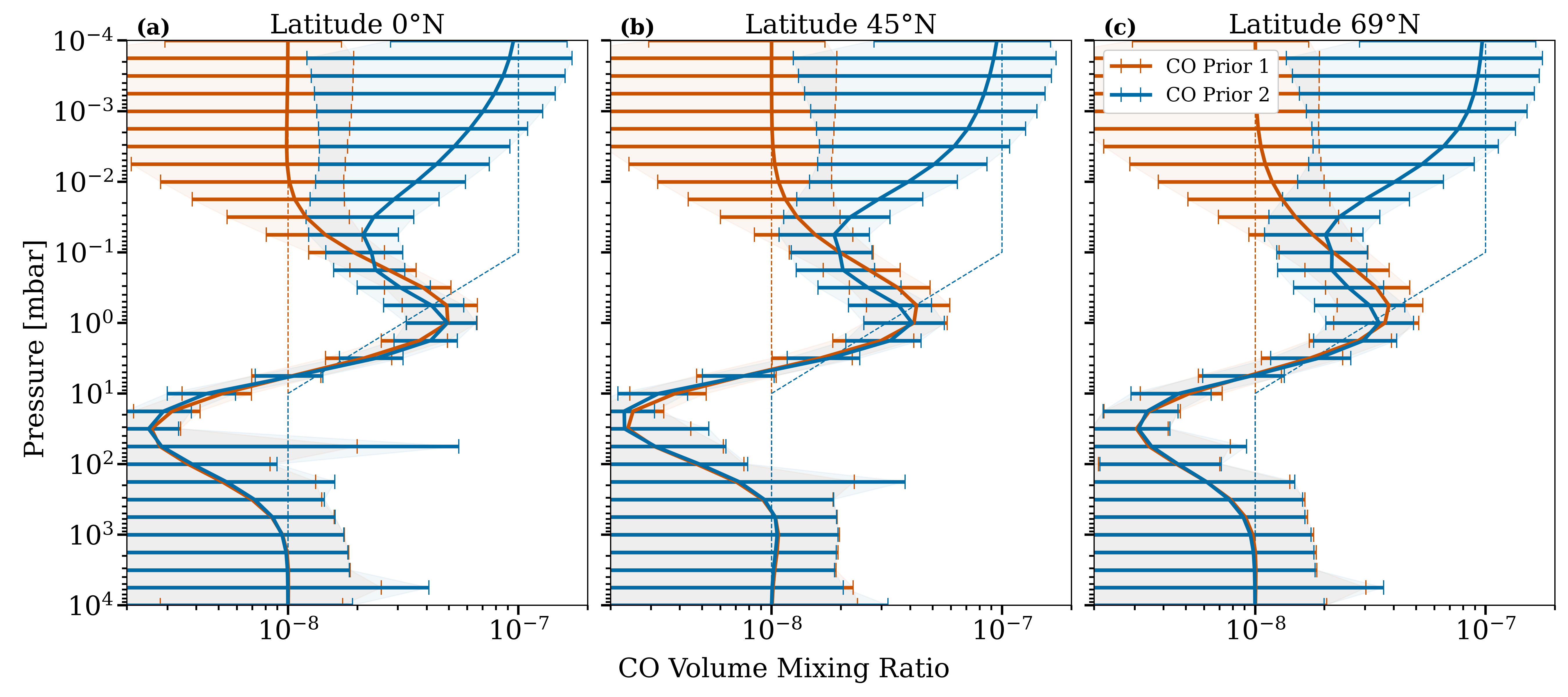}
   \caption{Vertical profiles of CO abundance obtained with our two retrieval tests at (a) the equator, (b) 45$^\circ$N and (c) 69$^\circ$N. Here, we have averaged the resulting vertical profiles from the two limbs. }
   \label{fig:comparison_profiles}
\end{figure*}

The resulting CO profiles at three latitudes (equator, 45$^\circ$N, and 69$^\circ$N) for both prior profiles tested here are presented in Figure \ref{fig:comparison_profiles}. 
For pressures lower than 0.1 mbar, each retrieved profile is pulled back to its prior,  with an increase in the errors, consistent with the absence of sensitivity for the highest altitudes in our modelled atmosphere. 
For the 0.1--1 mbar pressure range, to which the observations are most sensitive (Figures \ref{fig:contribution_function} and \ref{fig:app_contribution_function_prior2}), the resulting profiles show consistent abundance values between the two priors across all latitudes. 
For the 1--10 mbar pressure range, both profiles depict a decrease in CO abundance by 1 order of magnitude with increasing pressure, which is better constrained at 45$^{\circ}$N. 
Finally, resulting CO profiles are pulled back to their prior for pressures higher than 10 mbar, with large error-bars, typical of inversion calculations that lack observational constraints. 
Hereinafter, we will therefore focus on the CO abundances within the 0.1--10 mbar pressure range, which is the region of highest sensitivity of this ALMA dataset. 

\section{Results}
\label{sec:results}
\subsection{Meridional distribution of CO} 
\label{subsec:results_merid_CO}

\begin{figure}
    \centering
    \includegraphics[width=\linewidth]{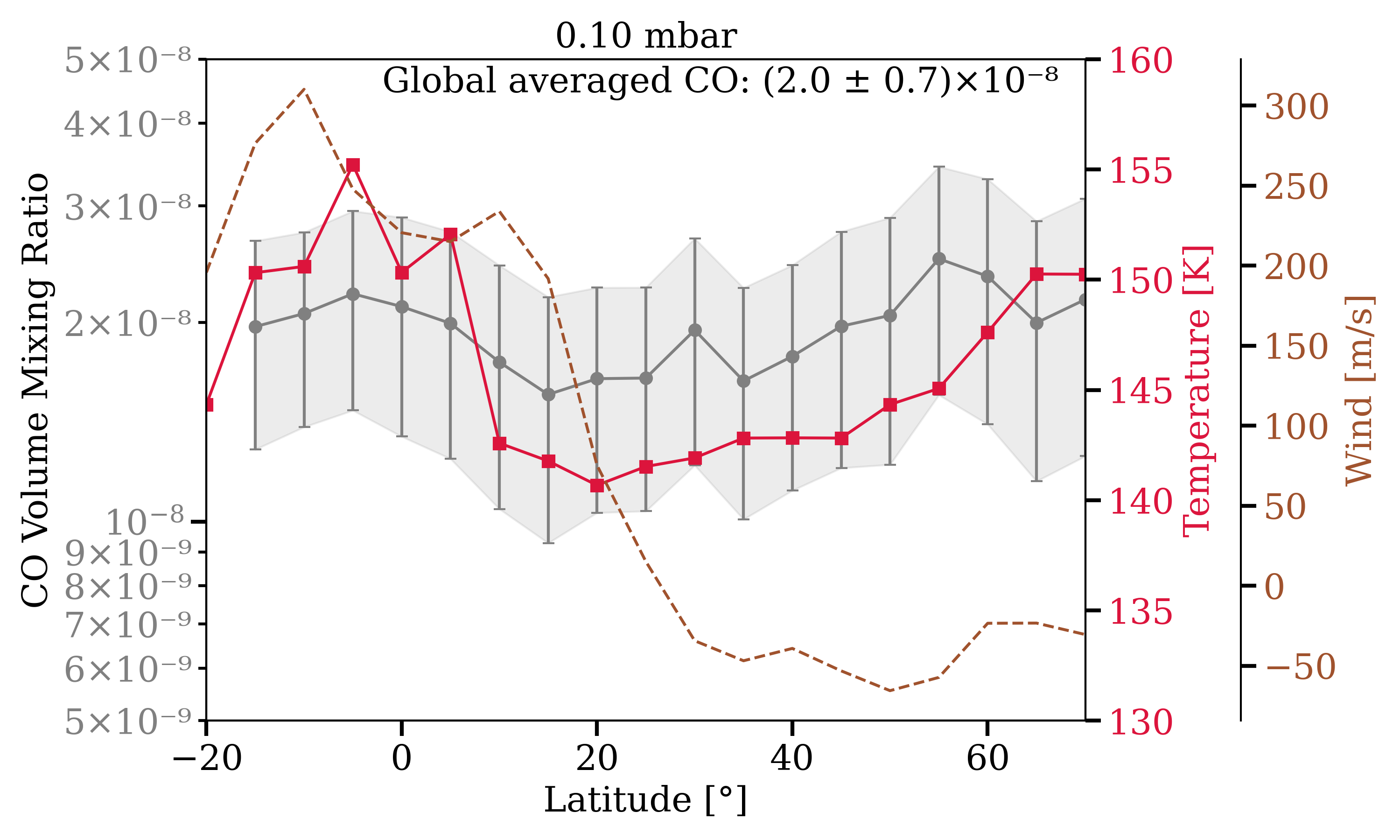}
    \caption{Meridional variations of CO abundance at 0.1 mbar, superimposed with the temperature at 0.1 mbar as retrieved from Cassini/CIRS in 2017 (red line, displayed in Figure \ref{fig:temperature_inputs}) and 
    the retrieved zonal wind from the same CO observations probed from 0.01 to the 20 mbar level \citep[brown dashed line,][]{Benm:22,Benm:25}.}
    \label{fig:CO_merid_01mbar}
\end{figure}

\begin{figure}
    \centering
    \includegraphics[width=\linewidth]{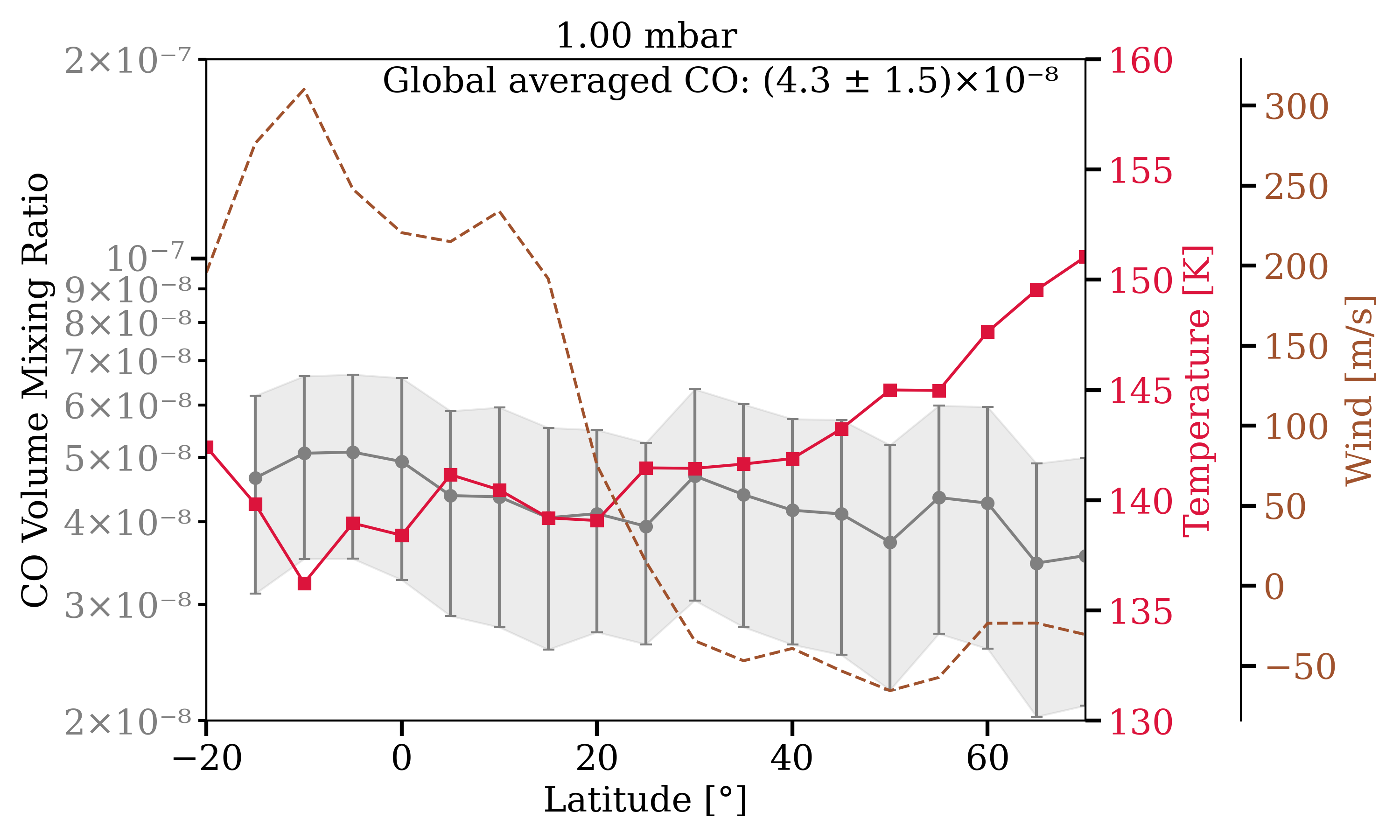}
    \caption{Same as Figure \ref{fig:CO_merid_01mbar} but at 1 mbar.}
    \label{fig:CO_merid_1mbar}
\end{figure}

Since CO is a long-lived species in Saturn's atmosphere, we can assume that its abundance is zonally homogeneous. 
Therefore, to reduce the error bars, we averaged the two meridional profiles obtained at the western and eastern limbs for each inversion, i.e., using Prior 1 or Prior 2. 
We have further smoothed the two meridional profiles over 5$^\circ$-bins, a latitudinal width consistent with the spatial resolution of the ALMA main array (see section \ref{sec:obs}). 
This step does not reduce the error bars, as adjacent beam pointings are not independent; rather, it shows measurements that could be considered independent of each other.
Examples of bin-averaging for each CO prior are presented in Figures \ref{fig:app_bin-av_prior1} and \ref{fig:app_bin-av_prior2}. 
In addition, due to the consistency of the CO prior profiles within the 0.1–1 mbar pressure range (see Figure \ref{fig:comparison_profiles}), we have averaged the two meridional distributions and calculated the mean of their error-bars to present our best estimate of the zonal-mean meridional profile of CO at 0.1 and 1 mbar in Figures \ref{fig:CO_merid_01mbar} and \ref{fig:CO_merid_1mbar}, respectively.
We also compare this mean meridional variation of CO abundance with the temperature field at the same pressure level used as input to the forward model (shown in Figure \ref{fig:temperature_inputs}), and 
the retrieved zonal winds deduced using our CO dataset by \cite{Benm:22,Benm:25}, which have been re-binned on a 5$^{\circ}$-bin-width latitude grid.

Overall, the meridional distribution of CO at both 0.1- and 1-mbar pressure levels is mainly homogeneous throughout the probed latitudinal range, with a mean value of $(2.0\pm0.7)\times10^{-8}$ at 0.1 mbar and a mean value of $(4.3\pm1.5)\times10^{-8}$ at 1 mbar (all abundances are given in volume mixing ratios). 
There are three marginal departures from the homogeneous meridional profile: (i) an enhancement of CO abundance at the equator compared to its surroundings ($\pm$20$^\circ$) observed at 0.1 mbar, (ii) a local enhancement at 30$^\circ$N observed at 0.1 mbar only, and (iii) another local enhancement close to 60$^\circ$N observed at 0.1 and 1 mbar, which will be discussed later in this section. 

The CO meridional abundance in the equatorial region ($\pm$20$^\circ$) shows the signature of a local enhancement at both 0.1. 
For one pressure level to the next, the equatorial enhancement varies: from $1.2\times10^{-8}$ at $\pm$20$^\circ$ to $2\times10^{-8}$ at 0$^\circ$.
During the Cassini Grand Finale mission, the INMS, CDA, and MIMI instruments observed a large amount of material entering at the top of Saturn's atmosphere, spread over a narrow latitude band of $\pm5^\circ$ centered on the equator. 
Considering that the downward transport of material from the top of the atmosphere down to 0.1–1 mbar occurs on a timescale of $\approx$10 years \citep{Mose:05saturn}, shorter than the timescale for meridional redistribution larger than a Saturnian year \citep{Bard:22}, such a large and permanent source would cause a sharp enhancement in CO around the equator, which seems inconsistent with the wide and weak enhancement resulting from our retrieval calculations.
This would suggest that the equatorial inflow observed by Cassini was, at the very least, too recent to be visible or overestimated).

% \begin{figure}
%     \centering
%     \includegraphics[width=\linewidth]{meridional_CO_mean_T_W_p_1.00e-02_single.png}
%     \caption{Same as Figure \ref{fig:CO_merid_01mbar} but at 10 mbar.}
%     \label{fig:CO_merid_10mbar}
% \end{figure}

Finally, our results show a possible local enhancement of CO between 50 and 60$^\circ$N from 1 mbar to 0.1 mbar. 
This enhancement remains marginal compared with the individual error bars. 
Compared to the temperature variation at this location, the retrieved CO abundance depicts a unique signature that cannot be obviously linked to the thermal field. 
Furthermore, it cannot be linked to any dynamical structure revealed by the wind measurements (Figures \ref{fig:CO_merid_01mbar} and \ref{fig:CO_merid_1mbar}).
The CO enhancement location is reminiscent of the 62$^\circ$S signature of the H$_2$O -- attributed to material originating from Enceladus plumes and transported along the magnetic field lines \citep{Moor:15SaturnRingRain,ODon:19SaturnRingRain}. 
It may therefore result from ionized water coming from Enceladus, as a northern counterpart to the southern H$_2$O influx. 
However, the inferred vertical CO profile at these latitudes decreases with pressure from 1 to 0.1 mbar, similarly to other latitudes. 
This behaviour implies that the chemical conversion from H$_2$O to CO would need to occur more efficiently at larger pressures than at lower ones. 
This hypothesis thus requires confirmation through photochemical modelling.

% \FloatBarrier
\subsection{Vertical distribution of CO} 
\label{subsec:results_verti_CO}

The prominent feature that stands out from our retrieval is the increase in abundance from a mean value of $(2.0\pm0.7)\times10^{-8}$ at 0.1 mbar to a mean value of $(4.3\pm1.5)\times10^{-8}$ at 1 mbar, followed by a decrease in abundance to a mean value of $(5.1\pm1.7)\times10^{-9}$ at 10 mbar. 
This vertical distribution stands out for each independent latitude at the 1 or 2-sigma level (Figures \ref{fig:CO_merid_01mbar}-\ref{fig:CO_merid_1mbar}). 
It is also a consistent feature of the inversion, using the two different priors (Figure \ref{fig:comparison_profiles}). 
We therefore consider this shape of the profile between 0.1 and 10 mbar to be robustly established by our dataset.

The increase in abundance with increasing pressure over 0.1-1 mbar contradicts most previous observations and photochemical modelling studies. 
For example, based on submillimeter observations of CO (J=6-5) with the James Clerk Maxwell Telescope (JCMT), \cite{Cava:10} found that in this pressure range (0.1-1 mbar), the abundance of CO is close to vertical homogeneity. 
Similarly, photochemical models, including different scenarios and/or values of external material inflow from a cometary impact alone or from a steady-source, all agree on vertical homogeneity or a slight decrease in CO abundance with pressure between 0.1 and 1 mbar \citep{Mose:00PartII,Mose:23ringrain}. 
Nevertheless, \cite{Mose:17} and \cite{Mose:23ringrain} tested scenarios accounting for the combined effects of several sources of exogenic material. 
In \cite{Mose:17}, the scenario that models a source of H$_2$O from  Enceladus flowing in at the top of the atmosphere, along with a 400 year-old cometary impact and a background injection of CO and H$_2$O molecules from the ablation of icy grains, results in a decreasing abundance of CO for pressures smaller than 0.3 mbar \citep[see Figure 11 in][]{Mose:17}. 
Their resulting CO profile predicts the maximum value of CO abundance at 0.3 mbar, while we predict the maximum value at 1 mbar. 
Their result also predicts a slower increase, with about a tenfold increase in abundance (from $10^{-8}$ at $0.6\times10^{-4}$ mbar to $9\times10^{-8}$ at 0.3 mbar) over more than a two-decade drop in pressure, whereas our results show an increase over a pressure drop of one decade (from 0.1 to 1 mbar). 
There is also a slower decrease from 0.3 mbar to 20 mbar for \cite{Mose:17}, whereas we show a decrease from $4.3\times10^{-8}$ at 1 mbar to $5.1\times10^{-9}$ at 10 mbar for our equatorial profiles in Figure \ref{fig:comparison_profiles}.
While the overall shape of \cite{Mose:17} and our CO profiles tends to confirm a cometary source scenario for the exogenic species inflow in Saturn's stratosphere, the quantitative contradiction in the vertical distribution of CO implies an inconsistency in either the age of the cometary impacts, or the amount of injected molecules for either CO or H$_2$O. 

However, at a pressure level of 1 mbar, the absolute values of CO abundance that we obtained are in agreement with previous observations as well as with modelling studies. 
On average, at 1 mbar, our retrieval calculations result in a CO abundance of $(4.3\pm1.5)\times10^{-8}$ vmr, consistent with \cite{Cava:10}, who estimated an abundance of  approximately 6$\times$10$^{-8}$ vmr from JCMT observations while considering a 220-year-old cometary impact as an external source. 
With a photochemical model including a source of water ejected from Enceladus, water and other oxygen-bearing species from the rings, and CO deposited from a large cometary impact, \cite{Mose:00PartII} and \cite{Mose:17} predicted a global CO abundance ranging from 10$^{-9}$ to 10$^{-7}$ vmr at 1 mbar, depending on the exogenous source scenario considered (different combinations of values for the above-mentioned sources).

\section{Assessing the origin of CO}
\label{sec:discussion_equatorial_origin}

\subsection{A permanent, large equatorial influx}
We first examine whether our observational results are compatible with a permanent, large equatorial influx of material falling from the rings, as observed during Cassini's Grand Finale \citep{Wait:18}. 
Two properties of our retrieved 2-dimensional CO vmr field can be used for this assessment.

First, the inverted vertical profiles only marginally differ between the equatorial and mid-latitudes. 
At all latitudes, the CO abundance peaks around 1 mbar, increases in abundance between 0.1 and 1~mbar, and then decreases in abundance between 1 and 10 mbar. 
This similarity between all latitudes indicates neither a specific nor a significant process occurring only at equatorial latitudes. 
Moreover, the negative abundance gradient between 1 and 0.1~mbar does not point toward a steady external flux that would, in turn, produce a positive vmr vertical gradient towards the upper atmosphere \citep{Cava:10,Mose:23ringrain}.
Second, our CO abundance distribution does not show any sharp enhancement in CO around the equator that would be the result of the large amounts of material entering the latitudinal band of $\pm$5$^\circ$ about the equator at the top of Saturn's atmosphere, as derived from the INMS, CDA, and MIMI instrument measurements during Cassini's Grand Finale. 
 
More quantitatively, we further attempt to derive an order-of-magnitude upper limit on the mass of the equatorial influx based on our measurements.
Our enhancement of $3\times10^{-8}$ on the CO abundance from $\pm$20$^\circ$ to 0$^\circ$ corresponds to an enhancement of $7.5\times10^{10}$ cm$^{-2}$ in the CO column density above the 1-mbar pressure level. 
Taking into account this equatorial enhancement between the equatorial region and the mid-latitudes, we can constrain the total additional mass of CO between the 5\degr S--5\degr N equatorial band to be $\leq 1.3\times10^{10}$~kg. 
Given that the combined CO+CO$_2$ influx has been measured to be on the order of $10^3$~kg/s \citep{Seri:22SaturnRings}, our upper limit on the mass implies that the duration of the inflow must have been shorter than $1.2\times 10^7$~seconds, i.e., 130 terrestrial days. 
This short timescale is comparable to Cassini's Grand finale, when proximal orbits probed the space between the rings and the atmosphere of Saturn (from 22 April 2017 to the final plunge on 15 September 2017). 
This remains an order-of-magnitude for several reasons. 
Firstly, it assumes that the incoming CO is in the gas phase, which is not certain according to the calculations of \cite{Mose:23ringrain}, suggesting a scenario with falling particles that are so small that ablation is not capable of chemically producing more CO. 
Secondly, a more reliable estimate of the equatorial influx should take into account the effect of meridional transport and the differences in the CO vertical distributions between the equatorial region and the mid-latitudes. 
Indeed, with the large equatorial influx, the CO vertical profile at equatorial latitudes should peak at low pressures outside the pressure range favoured by our inversion, resulting in slightly different estimates of the CO column density.

Our retrieval results thus reject the possibility of a stable and long-lived source of exogenous material originating from the rings at the equator (in the form of CO itself or material that can be chemically converted into CO). 
This conclusion is consistent with the chemical modelling study by \cite{Mose:23ringrain}.
Indeed, these authors proposed that the material detected during Cassini’s Grand Finale either resulted from an abnormal and transient infall event (whose impact would not yet be detectable at a pressure level of 0.1 to 1 mbar at the time of our ALMA observations) and/or was composed of small dust particles (less than 100 nm) that do not ablate. 
We draw the same conclusion with our HCN retrieval calculations \citep[see details in ][]{Fouc:26}. 

\begin{figure}
    \centering
    \includegraphics[width=\linewidth]{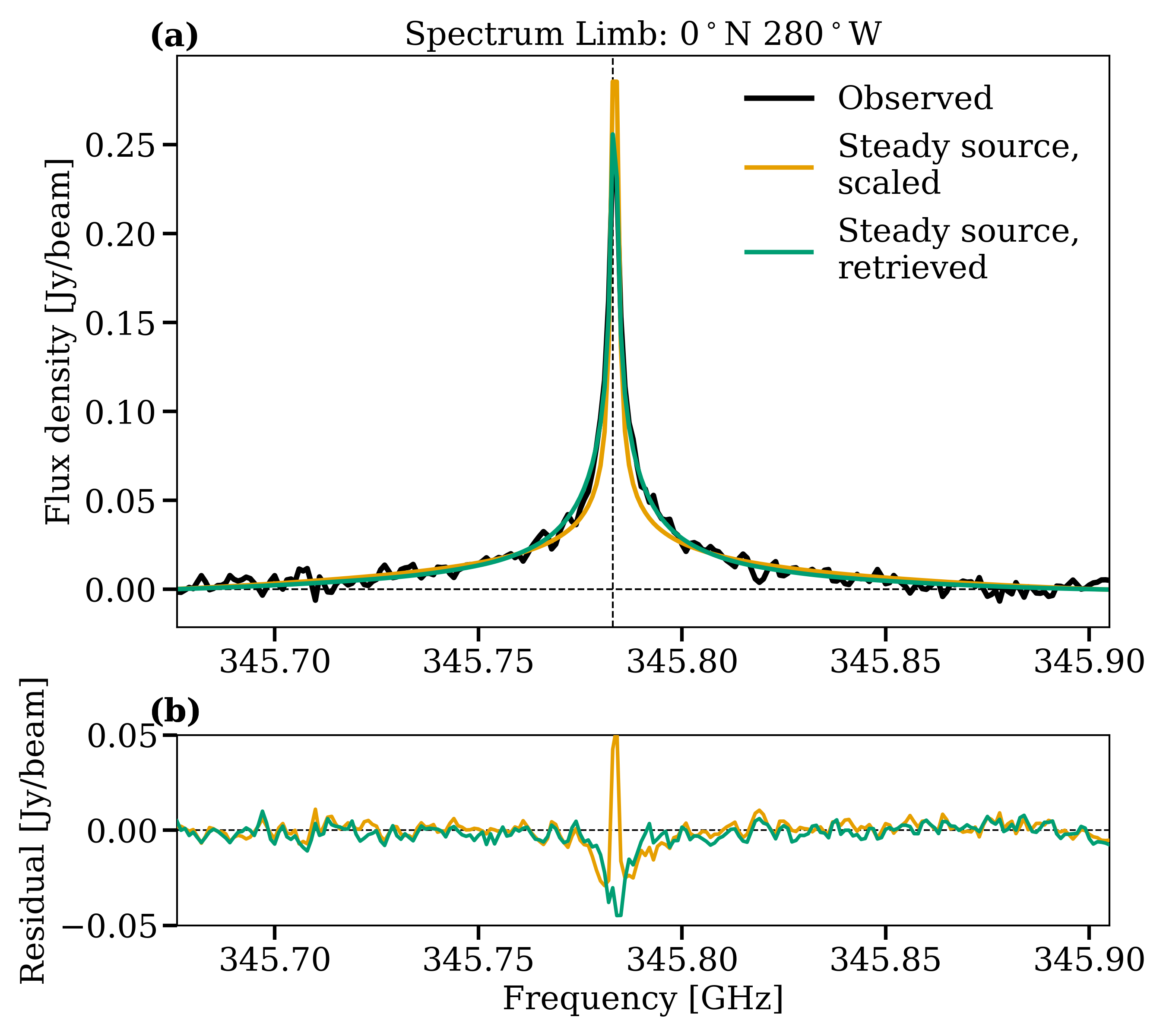}
    \caption{CO line fitting for (a) the equatorial eastern limb spectrum at 0$^\circ$N-280$^\circ$W for the \emph{retrieved} and \emph{scaled} methods in the case of a steady source of CO as prior profile (\emph{scaled} in yellow and \emph{retrieved} in green).
    (b) Residual flux density (observed spectrum - synthetic spectrum) for the observed limb spectrum.}
    \label{fig:spectrum_fitting_origins_steadysource}
\end{figure}
\begin{figure}
    \centering
    \includegraphics[width=\linewidth]{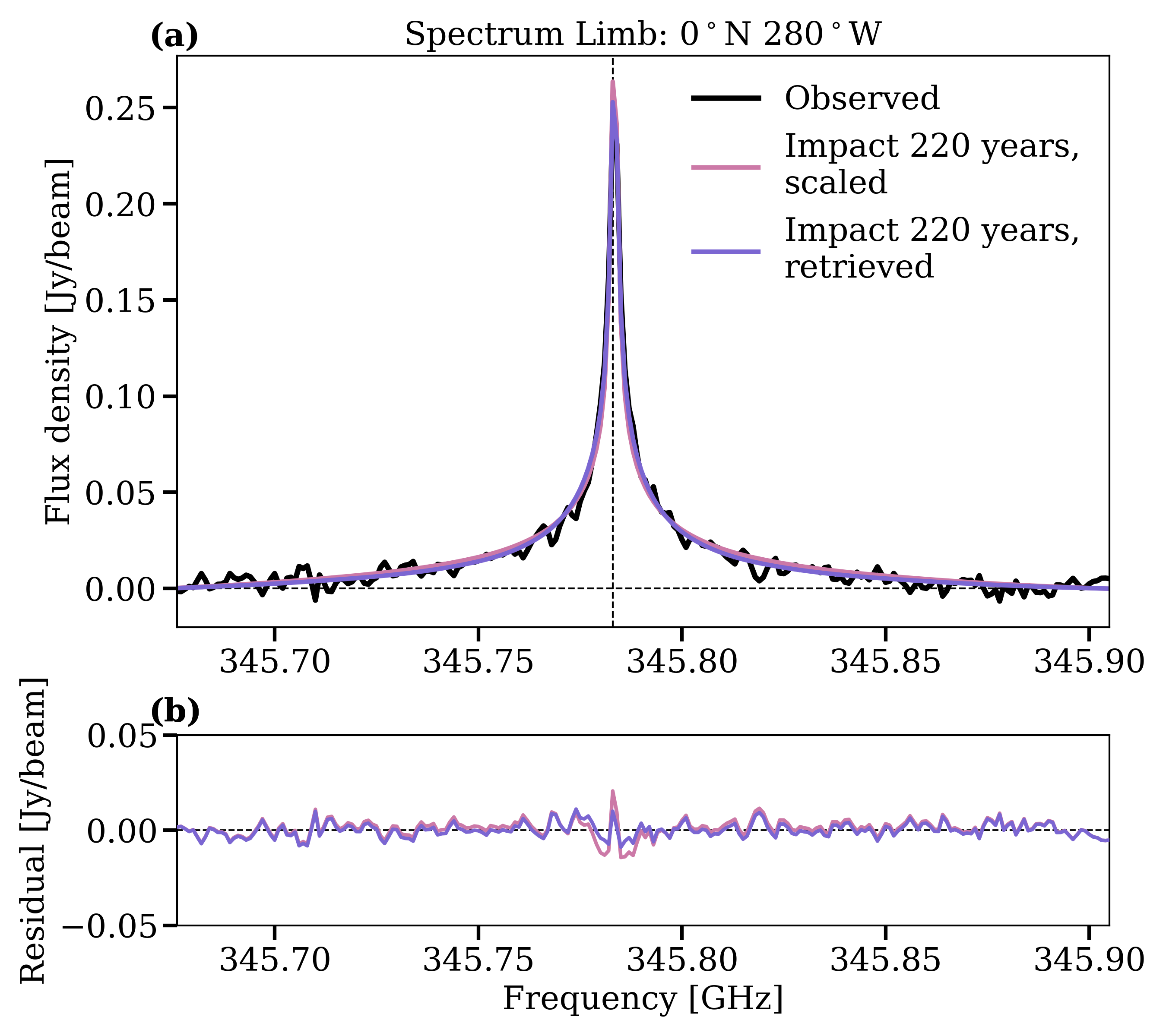}
    \caption{Same as Figure \ref{fig:spectrum_fitting_origins_steadysource} but in the case of a 220-year-old cometary impact as prior profile (\emph{scaled} in pink and \emph{retrieved} in purple).}
    \label{fig:spectrum_fitting_origins_comet220yo}
\end{figure}
\begin{figure*}
    \centering
    \includegraphics[width=0.9\linewidth]{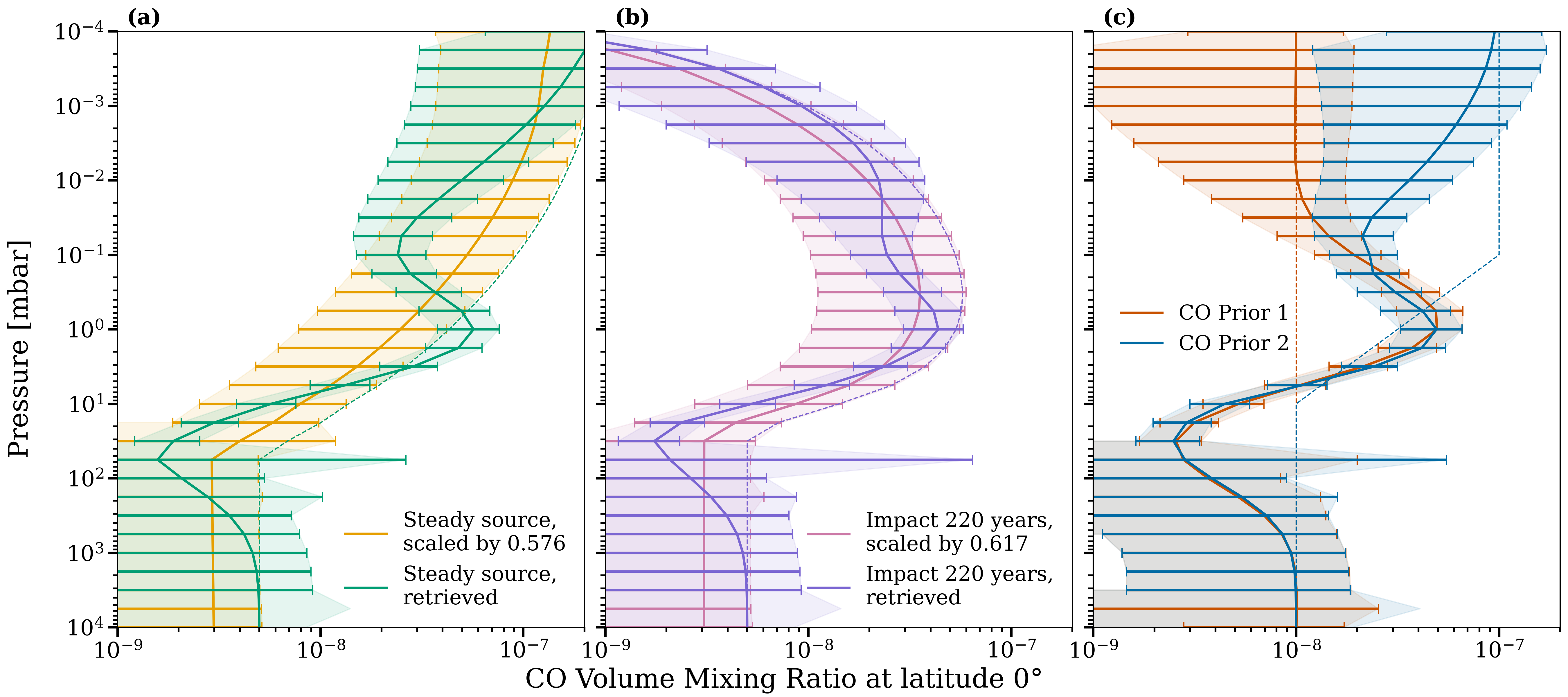}
    \caption{Retrieved CO profiles at the equator for the \emph{retrieved} and \emph{scaled} methods in the case of (a) a steady source of CO as prior profile (\emph{scaled} in yellow and \emph{retrieved} in green), (b) a 220-year-old cometary impact as CO prior profile (\emph{scaled} in pink and \emph{retrieved} in purple), and (c) the two initial spectral inversion scenario tested in this study (only \emph{retrieved}). Dashed lines represent the prior profiles.}
    \label{fig:CO_vert_origins}
\end{figure*}

\subsection{A weak, continuous, latitudinally extended external flux}
We next examined the possibility that Saturn's stratospheric CO originates from a weak, continuous, latitudinally extended external flux.
To test this hypothesis, we specifically used, as a prior profile in our inversion, a CO vertical profile proposed by photochemical models to represent the effect of a planetary-wide steady source. 
We adopted the steady source profile corresponding to a flux of $\phi_0=4.1\times10^6$~cm$^{-2}$ s$^{-1}$ that best matched previous CO observations in \cite{Cava:10} study, and performed the inversion using two different smoothing length scales: one equal to 1.5 times the atmospheric scale height ($H$), and a second equal to 50 scale heights, the latter being equivalent to scaling the whole prior profile by a constant factor. 
The spectral fits at the equator are shown in Figures \ref{fig:spectrum_fitting_origins_steadysource} and \ref{fig:spectrum_fitting_origins_comet220yo}, while the inverted profiles are presented in Figure \ref{fig:CO_vert_origins}. 
It is clear that the steady-state profile, even when scaled by a factor of about 0.7 relative to the prior, fails to reproduce the observed spectra. 
Its positive gradient with altitude overestimates the emission in line core and underestimates it in the line wings. 
For this reason, the inversion performed with a smaller restrictive smoothing constraint alters the prior between 1 and 0.1 mbar, converging toward a solution presenting a negative abundance gradient between these two pressure levels. 
This result is consistent with the conclusions drawn earlier (Section~\ref{SecVerticalResults}).

\subsection{A cometary impact}
Since the steady-state external influx scenarios, involving either a strong and concentrated flux at the equator or a weaker flux homogeneous over the whole planet, are not compatible with our dataset, we next investigated how the cometary impact scenario proposed by \cite{Cava:10} fares against our observations. 
In this objective, we performed two additional inversions on some selected spectra, adopting the 220-year impact profiles of \cite{Cava:10} as a prior and using the two different vertical correlation lengths of $1.5\times H$ (next labelled \textit{retrieved}) and $50\times H$ (\textit{scaled}).

Both the \emph{retrieved} and \emph{scaled} methods (shown as purple and pink lines on Figures \ref{fig:spectrum_fitting_origins_comet220yo} and \ref{fig:CO_vert_origins}) satisfactorily fit the peak of the equatorial CO line within the noise level of the spectra. 
However, the best agreement in the wings remains obtained by the \emph{retrieved} method, which implies a local depletion of CO at 0.1~mbar. 
This result confines the CO into a thin layer between 0.1 mbar and 10 mbar, yielding a vertical structure similar to that obtained with prior 1 and prior 2. 
The difference in CO variation between 0.1 and 1 mbar in the \emph{scaled} and \emph{retrieved} profiles may reflect sensitivity to the parameters of cometary impact models and might be resolved by modifying the turbulent diffusion parameters assumed in those models. 
The impact age is primarily inferred from the pressure level of the cut-off below which the CO abundance starts to decrease with increasing pressure. 
This cut-off corresponds to the location of the diffusion front of the material deposited during the impact. 
While our inversions are compatible with the 220-year impact proposed by \citet{Cava:10}, which rules out an older impact, whose deposited materials would have diffused deeper into the stratosphere, we cannot exclude a younger impact.
Additional modelling work on cometary impact diffusion is required to tightly constrain the age of the cometary impact.

\subsection{A high altitude loss of CO}
The results for either a continuous source or a cometary impact (in \emph{scaled} method) fail to reproduce the shape of the profile we obtained with this ALMA dataset. 
To explain the loss of CO at high altitudes, one can consider a chemical loss process tied to H$_2$O.
When photolyzed, H$_2$O yields OH, which reacts with CO to form CO$_2$, as discussed by \cite{Lell:02} for Jupiter and the impact of SL9. 
For Saturn, there are additional sources of water: Enceladus and the rings, which supply extra water to the upper atmosphere. 
Assuming the CO originates from a comet, it is deposited at around 0.1 mbar and then rapidly diffuses upwards \citep{Mose:17}. 
Therefore, in addition to the water resulting from the impact, the CO encounters the additional water concentrated at high altitudes. 
The reaction partially destroys it, and crucially, unlike SL9, where an equilibrium between CO and cometary H$_2$O is established, CO must constantly contend with water that keeps falling. 
This may lead to a CO profile with a decrease in its abundance towards higher altitudes. 

Conversely, the water consumed in the reaction must result in a
fairly steep rising profile of H$_2$O. 
We have some indications of this: \cite{Cava:19SaturnH2O} had to locate H$_2$O above the 0.3 mbar level in order to fit the Herschel data at both low and high spectral resolutions simultaneously. 
In addition, \cite{Lefo:25} showed that the 2011 beacon \citep[aftermath of the 2010 Great White Storm on Saturn,][]{Flet:12beacon} increased the water column by a factor of 7, a large enhancement possibly due to the sublimation of ice or vertical advection of a profile with a gradient.
Concerning CO$_2$, a large enhancement of abundance should be observed at a different pressure level than our maximum of CO vmr. 
However, the vertical resolution of the latest observations of CO$_2$ from the James Webb Space Telescope \citep{Flet:23} only allows for an estimated value of $(0.60 \pm 0.05)\times 10^{-9}$ at 0.5 mbar, homogeneous in latitude, consistent with previous observations from the Composite InfraRed Spectrometer on Cassini \citep{Abba:13}.
Moreover, modelling work from \cite{Mose:17} (Fig. 11) predicts a maximum $\mathrm{CO_2}$ concentration of around 0.2 mbar, which, in their case, is close to their resulting maximum CO concentration.

\section{Conclusions}
\label{sec:ccl}

In this study, we have performed CO abundance retrievals using ALMA observations of Saturn's stratospheric CO (J=3-2) line taken on May 25$^\mathrm{th}$ 2018, only a few months after the end of the Cassini mission. 
These data were initially used to provide the first absolute wind measurement in Saturn's stratosphere \citep{Benm:22,Benm:25}. 
This dataset covers the entire limb of Saturn's summer northern hemisphere with an unprecedented latitudinal resolution of 5$^\circ$.

Our dataset is mainly sensitive to the 0.1--10~mbar pressure range. 
Latitudinally-averaged at the 0.1, 1, and 10 mbar pressure levels, the mean CO abundance increases from $(2.0\pm0.7)\times10^{-8}$ at 0.1 mbar to a mean value of $(4.3\pm1.5)\times10^{-8}$ at 1 mbar, and decreases to $(5.1\pm1.7)\times10^{-9}$ at 10 mbar.
While this vertical structure is not consistent with what was inferred in most of the previous observational and modelling studies, the absolute CO abundance retrieved at 1 mbar remains consistent with the values reported in the literature \citep{Cava:09,Cava:10}. 

Both the meridional homogeneity and the vertical structure of the CO abundance field obtained in this study allow us to rule out some possible sources for Saturn's stratospheric carbon monoxide.
The absence of any pronounced enhancement around the equator in our dataset, and the negative vertical abundance gradient between 1 and 0.1~mbar support the conclusion of \cite{Mose:23ringrain} that the material falling from the rings detected during Cassini's final orbits resulted from a relatively recent event, likely less than a decade old. 
As a consequence, its impact on stratospheric chemistry would not yet be detectable in the 0.1–1 mbar pressure range sampled by our observations. 
Alternately, as proposed by \citet{Mose:23ringrain}, the particle sizes in this infalling material could be so small that ablation is inefficient, preventing the production of additional CO even if the influx was older.

Moreover, the vertical structure of the retrieved CO abundance field is inconsistent with a steady-state, spatially extended, continuous external source. 
Such sources are not expected to produce a negative abundance gradient between 1 and 0.1~mbar. 
Hence, our observations do not support the ring rain or Enceladus sources alone as the dominant contributors to Saturn’s stratospheric CO.
The profile resulting from a ~220-year-old comet impact, modeled only with vertical transport by \cite{Cava:10}, better matches the ALMA data, but it is not fully satisfactory. 
While the latitudinal redistribution obtained here is analogous to the fate of the exogenous materials delivered by the SL9 comet to Jupiter in 1994 \citep{More:03,Cava:23}, there are disagreements between the impact scenario and our data, in the form of the negative abundance gradient between 1 and 0.1 mbar that is not predicted by models \citep{Cava:10}. 

In line with the modelling study by \cite{Mose:17}, we confirm, using our ALMA data, that the vertical distribution of CO results from a combination of exogenous materials: a cometary impact, which causes the vertical diffusion of CO, and the injection of additional water from the rings and/or Enceladus, which, through photolysis, converts CO into CO$_2$ at high altitudes, around 0.1 mbar. 
However, further observations sensitive to different pressure ranges are needed to properly characterise the chemical reaction pathways between CO, CO$_2$, and H$_2$O, and to strengthen our understanding of photochemistry in Saturn’s stratosphere.
This could be achieved for Jupiter with the Submillimeter Wave Instrument \citep[SWI,][]{Hart:26JUICE} onboard Juice, which will probe H$_2$O and CO for p $>$ 0.01 mbar in limb viewing.

Additional ALMA observations of Saturn's stratospheric carbon monoxide would also be valuable. 
In particular, they are needed to assess the temporal evolution of its vertical and meridional distribution. 
Owing to the observing geometry of the present dataset, only the summer northern hemisphere was visible, and comparable CO observations with similar spectral resolution and spatial coverage are still lacking for the southern hemisphere. 
Moreover, we are now approaching the 10-year anniversary of the end of the Cassini mission, and most importantly for this study, the 10-year anniversary of the detection of ring material infalling into Saturn's equatorial atmosphere. 
At this stage, it is reasonable to expect that the potential effects of this influx on the meridional and vertical distribution of CO, HCN, H$_2$O, CO$_2$, and, more generally, photochemically active species may now be detectable.

\begin{acknowledgements}
    Bardet was supported by a postdoctoral fellowship in astrophysics at Paris Observatory-PSL. 
    Cavalié and Lefour were supported by the Programme National de Planétologie (PNP) of CNRS/INSU and by the Centre National d’Études Spatiales (CNES) France (ROR: https://ror.org/04h1h0y33).
    This paper makes use of the following ALMA data: ADS/JAO.ALMA$\#$2017.1.00636.S. 
    ALMA is a partnership of ESO (representing its member states), NSF (USA), and NINS (Japan), together with NRC (Canada), NSTC and ASIAA (Taiwan), and KASI (Republic of Korea), in cooperation with the Republic of Chile. 
    The Joint ALMA Observatory is operated by ESO, AUI/NRAO and NAOJ.
    This project was provided with computing HPC and storage resources by GENCI at CINES thanks to the grant 2025-A0180110391 on the Adastra's GENOA partition.
\end{acknowledgements}

\bibliographystyle{aa} % style aa.bst
\bibliography{newfred} % your references Yourfile.bib

\begin{appendix}
\section{Retrieved CO profiles comparison between the narrow and broad spectral windows}

\begin{figure}
    \centering
    \includegraphics[width=\linewidth]{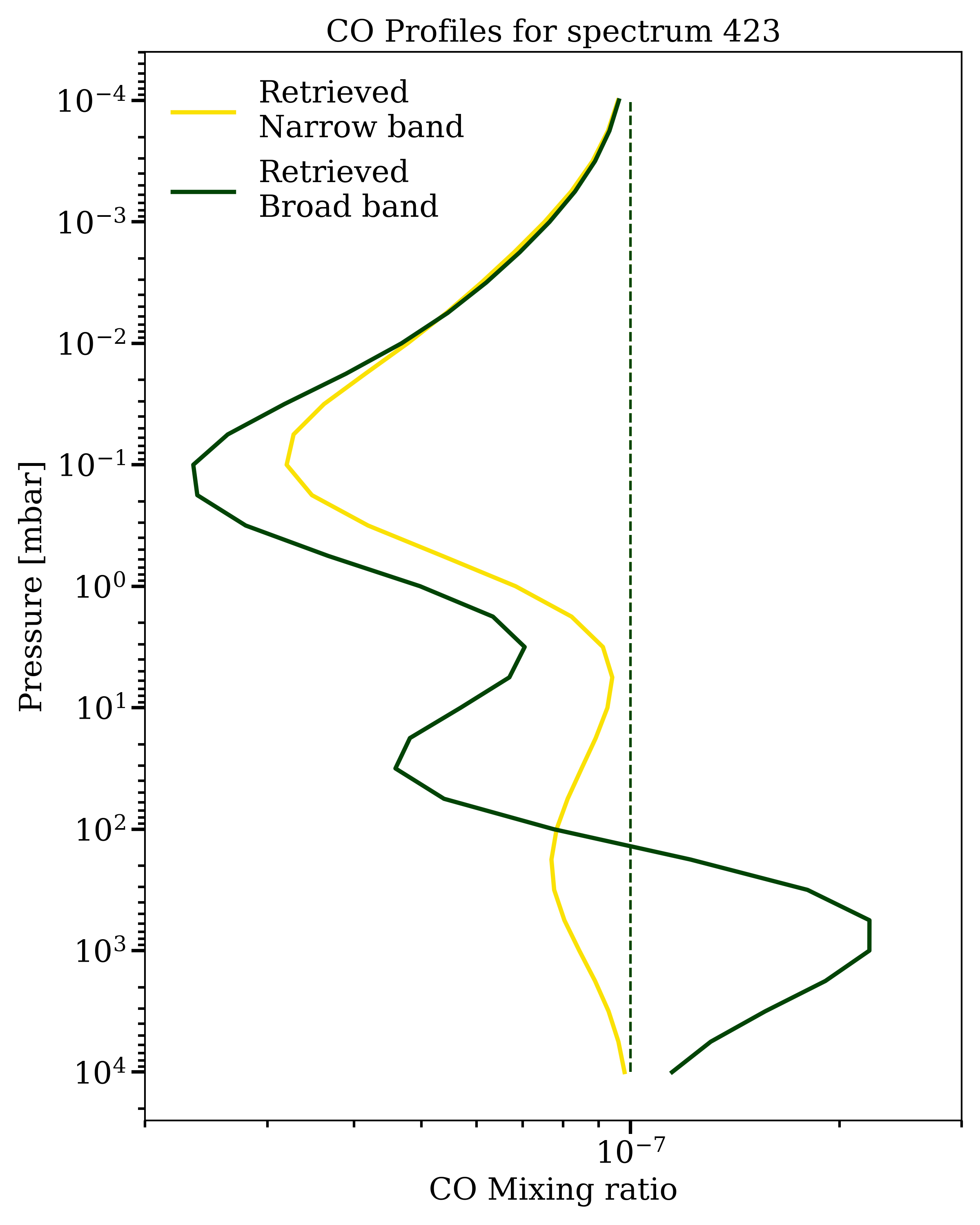}
    \caption{Comparison of retrieved CO profiles obtained for the western limb at 0$^\circ$N 280$^\circ$W using either the narrow band dataset (yellow) or the broad band (dark green). The arbitrary prior profile is depicted in dotted line.}
    \label{fig:app_compare_high-low_resolution_profiles}
\end{figure}

For both HCN (J=5-4) and CO (J=3-2), the $\#$2017.1.00636.S dataset is composed of two spectral windows centred on the rest frequency of the line: a narrow spectral window of 0.5 GHz (for HCN) and 0.120 GHz (for CO) in width (named hereafter narrow-band), and a large window of 2.0 GHz in width (named hereafter broad-band).  
Both narrow and large spectral windows are sampled with 2048 channels, implying a spectral resolution of 70 kHz for the narrow window and a spectral resolution of 1.1 MHz for the large one. 
Figure \ref{fig:app_compare_high-low_resolution_profiles} compares the retrieved vertical CO mixing ratio profiles obtained from the narrow-band and broad-band retrievals at 0$^\circ$N, 280$^\circ$W. 
The dashed vertical line indicates the a priori CO abundance of $10^{-7}$ volume mixing ratio.
Both retrievals show sub-solar CO abundances throughout the atmosphere, with values generally below the a priori profile. 

Between 0.1 and 10 mbar, the retrieved profiles diverge significantly. 
The broad-band retrieval develops two pronounced minima near 0.1 and 10 mbar, reaching abundances of only a few $2\times10^{-8}$ and $~10^{-9}$, respectively. 
The narrow-band retrieval maintains substantially higher and nearly homogeneous CO abundances; consequently, the maximum CO abundance at 1 mbar of $(5-6)\times10^{-8}$ is more easily noticeable and constrained by the broad-band retrieval. 
More importantly, the cut-off of the CO abundance profile (pressure between 1 and 10 mbar) is better assessed with the broad-band dataset, which is crucial information for the vertical distribution of CO in Saturn's stratosphere to determine the age of the cometary impact responsible for the injection of carbon monoxide and the resulting vertical diffusion of CO. 

\begin{figure}
    \centering
    \includegraphics[width=\linewidth]{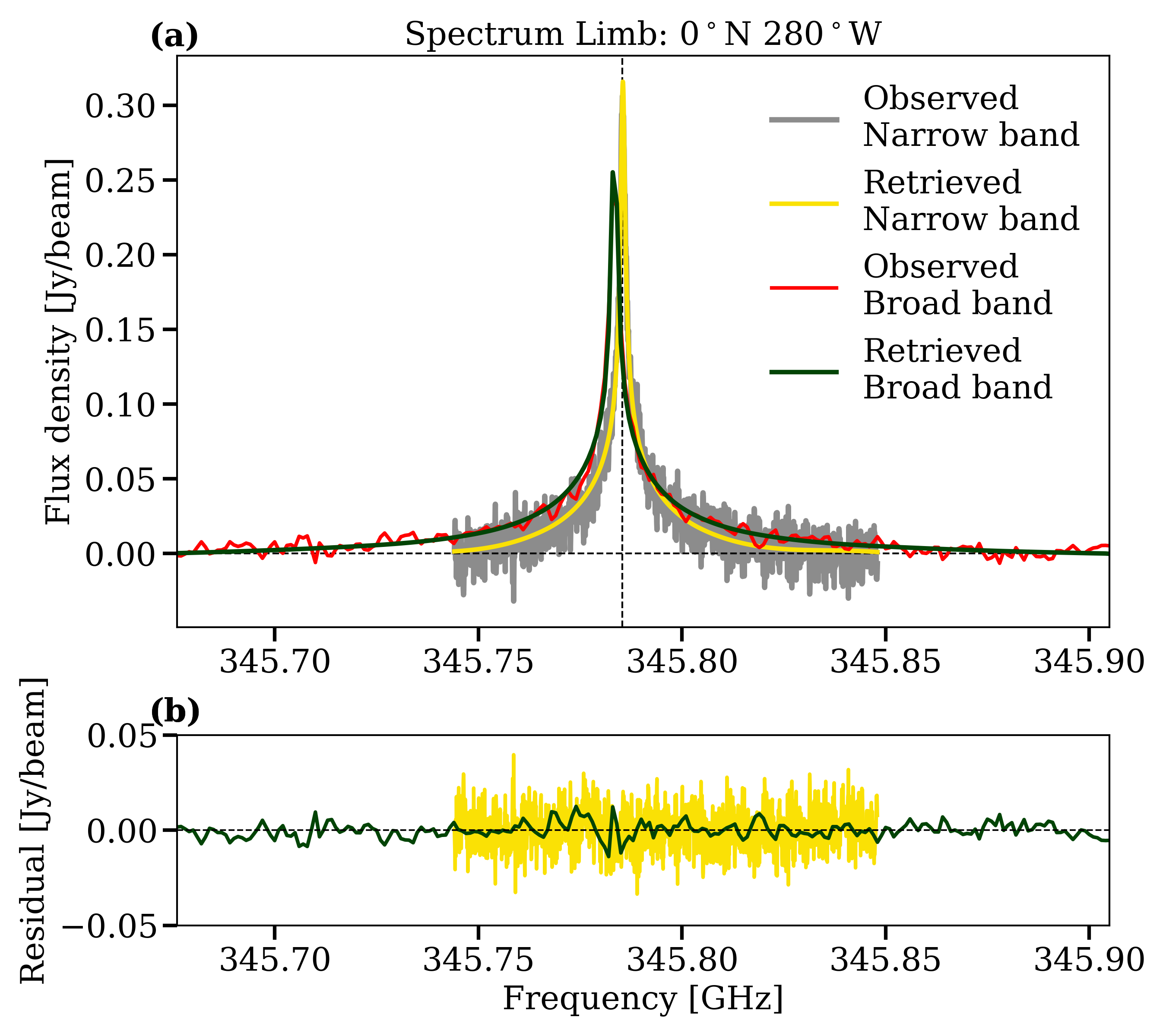}
    \caption{(a) Comparison of the CO line fitting obtained for the western limb at 0$^\circ$N 280$^\circ$W using either the narrow band dataset (yellow) or the broad band (dark green). (b) Residual flux density (observed spectrum - synthetic spectrum) showing that for both bands, the arbitrary prior profile in dotted line on Figure \ref{fig:app_compare_high-low_resolution_profiles} is fitting the observed limb spectrum within the noise of the observations.}
    \label{fig:app_compare_high-low_resolution_spectrum}
\end{figure}

Above approximately $10^{-2}$ mbar, the two solutions are relatively similar, exhibiting a gradual increase in CO abundance with pressure from $~10^{-7}$ to $(2-4)\times10^{-8}$, which shows that the retrieved profiles revert to the prior profile, characteristic of a lack of information in the spectra. 
In the same manner, below 10 mbar, the narrow-band retrieved profile depicts a gradual decrease of CO with pressure, ultimately reaching the prior profile abundance at the bottommost pressure of our modelled atmosphere (typical of a lack of information content in the spectra). 
Below 10 mbar, the broad-band solution exhibits large excursions, including an increase above the a priori abundance near 10$^{3}$ mbar. 
These large oscillations outside the sensitive pressure range (here 0.1 to 10 mbar) are also a consequence of a lack of information; therefore, the maximum abundance reached at 10$^{3}$ mbar is not a conclusive result. 

\begin{figure}
    \centering
    \includegraphics[width=\linewidth]{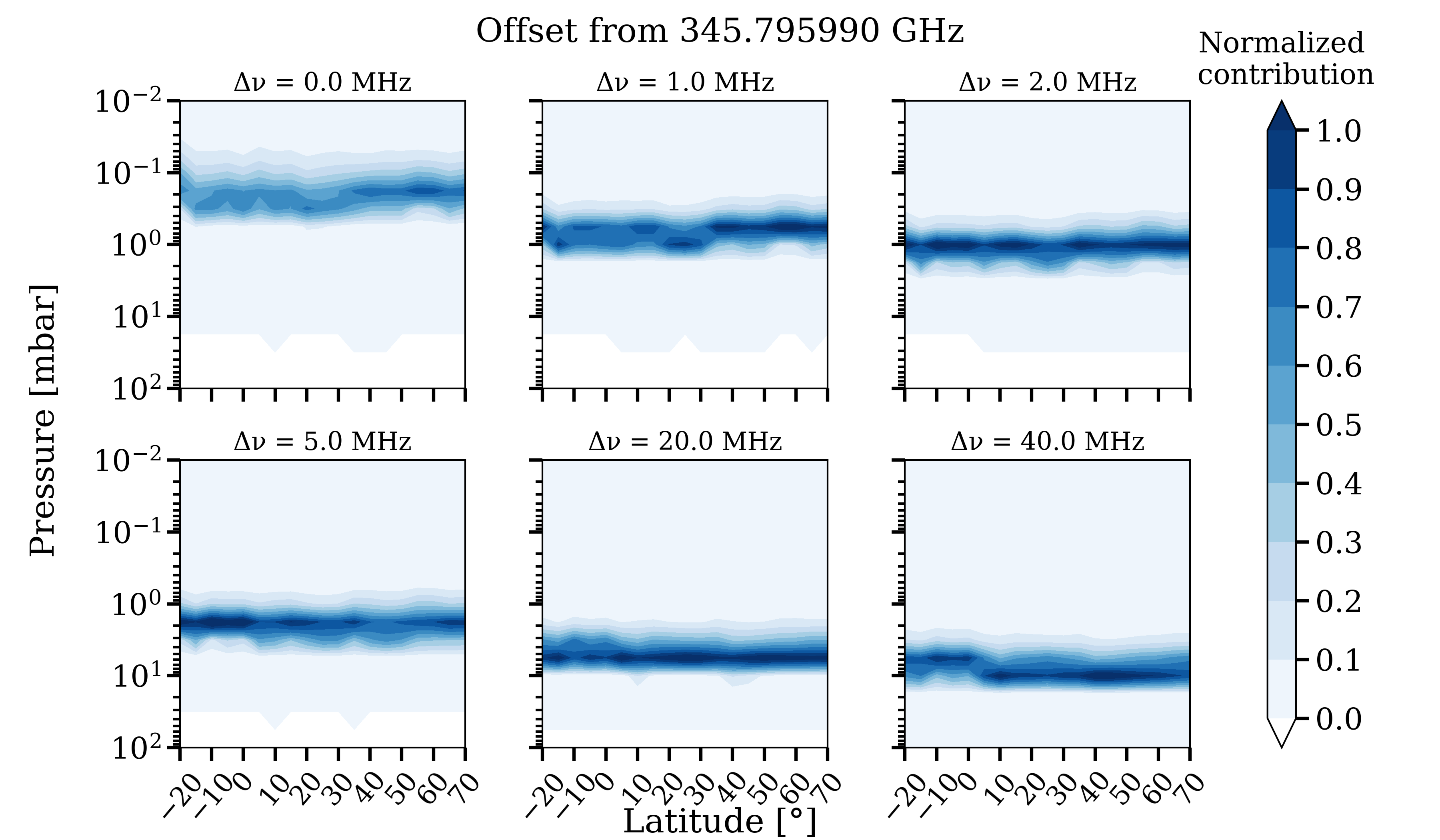}
    \caption{Normalize contribution function for the narrowband retrieval test in Figures \ref{fig:app_compare_high-low_resolution_profiles} and \ref{fig:app_compare_high-low_resolution_spectrum}.}
    \label{fig:app_contrib_narrowband}
\end{figure}

\begin{figure}
    \centering
    \includegraphics[width=\linewidth]{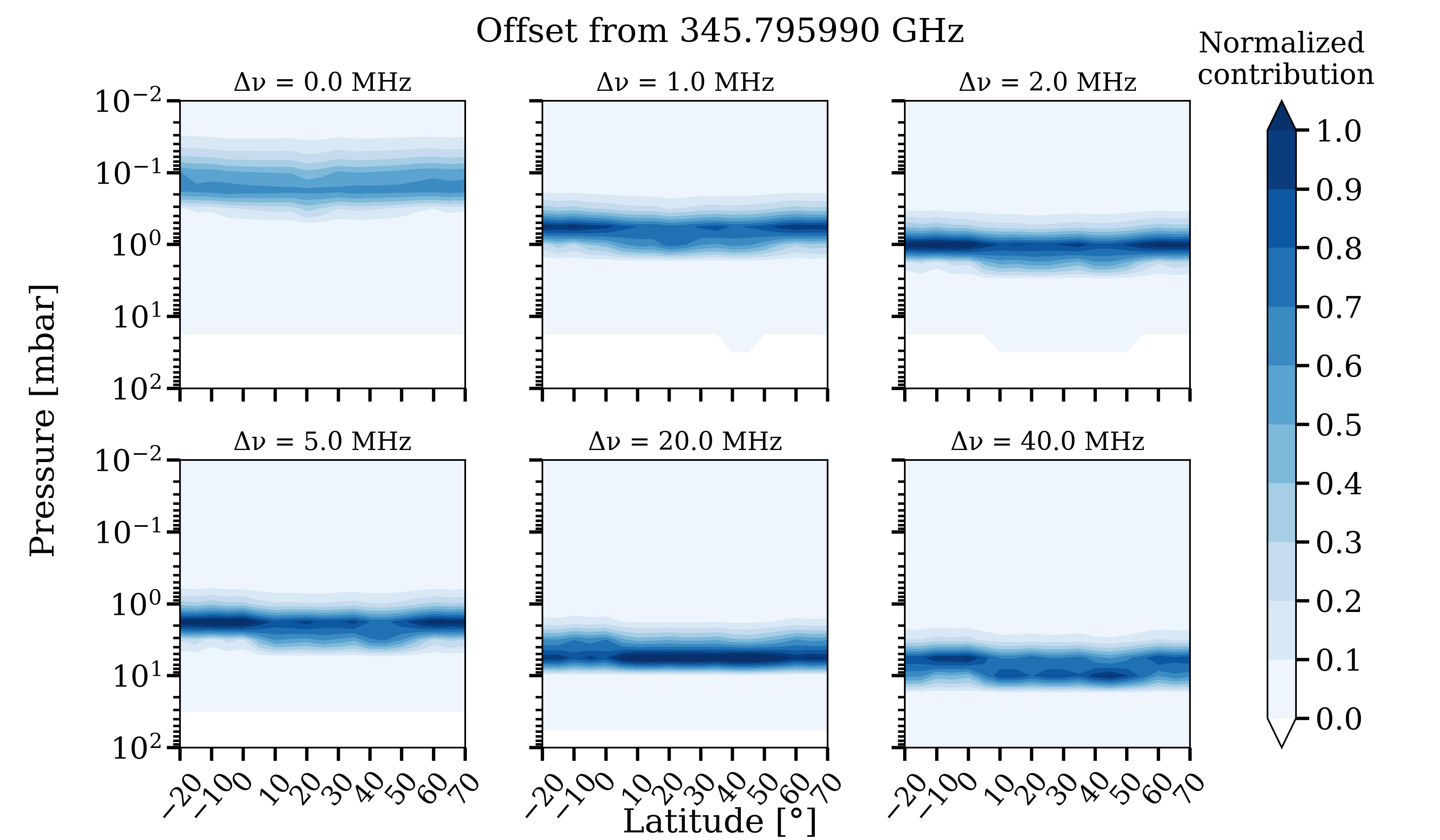}
    \caption{Normalize contribution function for the broadband retrieval test in Figures \ref{fig:app_compare_high-low_resolution_profiles} and \ref{fig:app_compare_high-low_resolution_spectrum}.}
    \label{fig:app_contrib_broadband}
\end{figure}

Comparison between the narrow-band, high-spectral-resolution spectrum and the broad-band, low-spectral-resolution spectrum reveals a loss of intensity in the CO line.
However, this loss of signal does not affect the vertical accuracy of our estimate.
Indeed, the contribution functions for these two reconstruction calculations (Figures \ref{fig:app_contrib_narrowband} and \ref{fig:app_contrib_broadband}) show equivalent sensitivity at the same pressure levels.

\onecolumn

\twocolumn

\section{CO line fitting at additional latitude points}
\label{app:line_fitting}

\begin{figure}[h!]
    \centering
    \includegraphics[width=\linewidth]{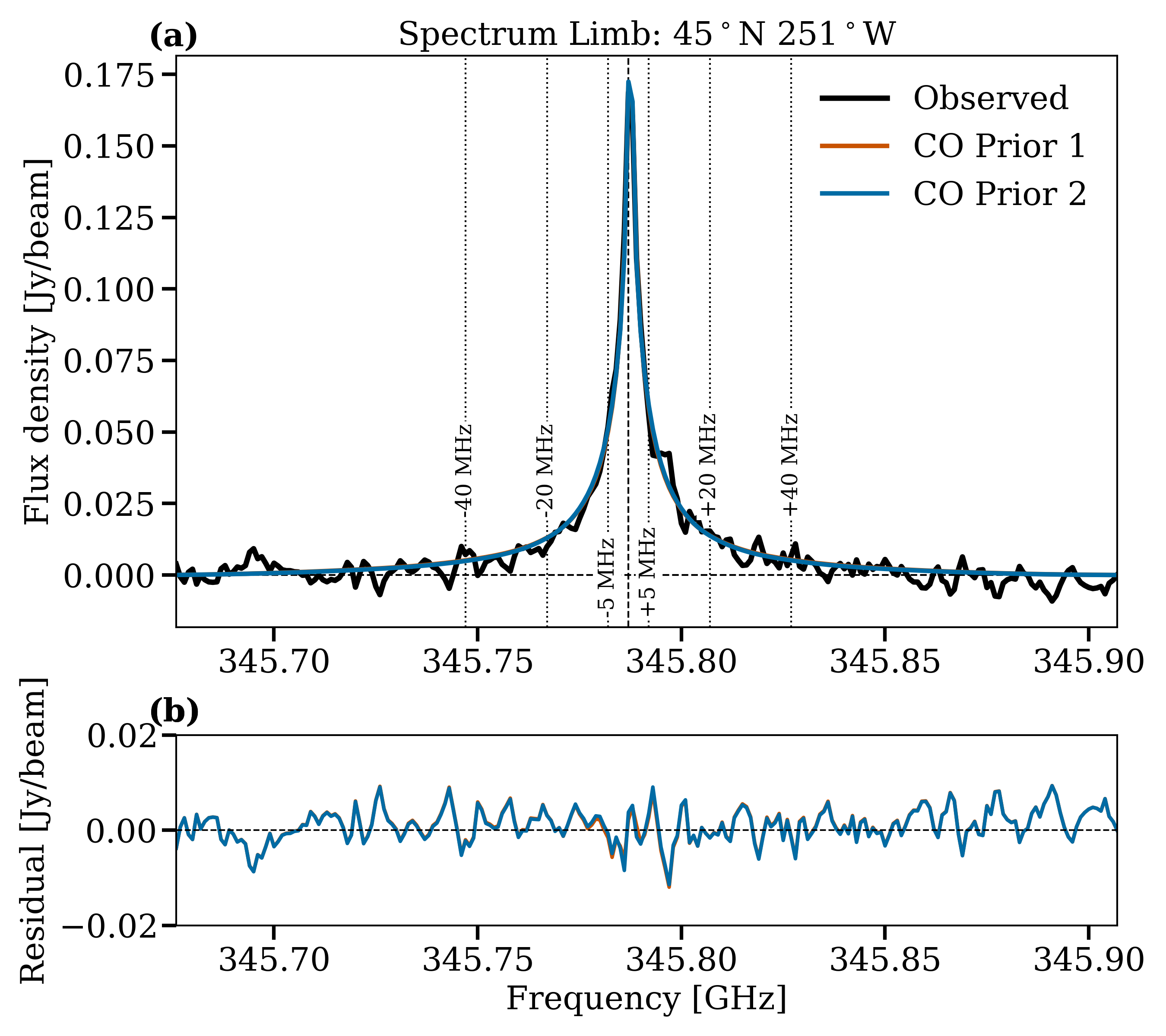}
    \caption{Example of the CO line fitting  for (a) the eastern mid-latitude limb spectrum (at 45$^{\circ}$N-251$^\circ$W) for the two retrieval calculations carried in this study, and (b) the residual flux density (observed spectrum - synthetic spectrum) showing that all CO prior profiles are fitting the observed limb spectrum within the noise of the observations. The black dashed line indicates the frequency of the peak, and the dotted lines depict different values of frequency offset from the peak, for which we have calculated the contribution function.}
    \label{fig:app-line_fitting_eastern45}
\end{figure}

\begin{figure}
    \centering
    \includegraphics[width=\linewidth]{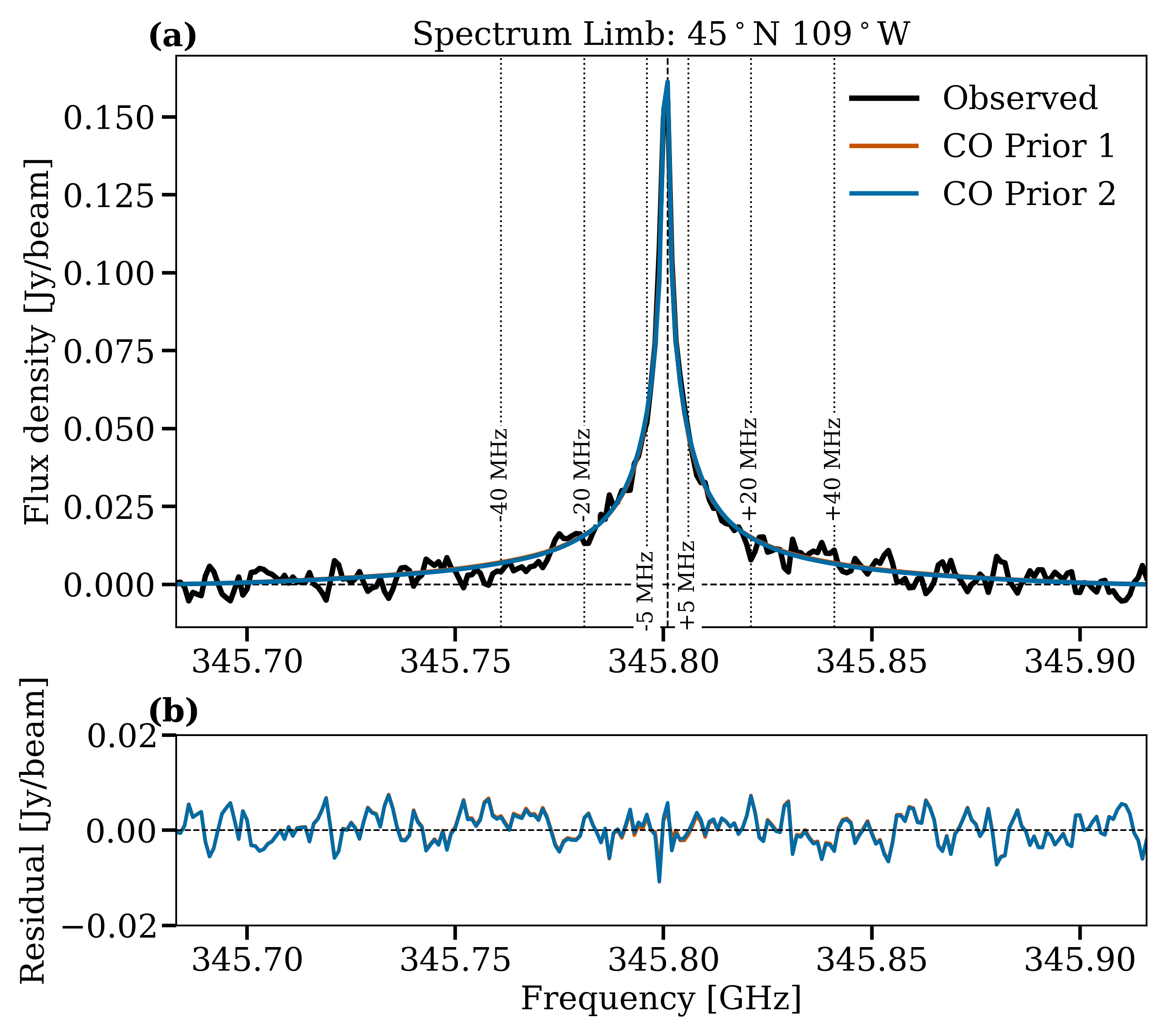}
    \caption{Same as Figure \ref{fig:app-line_fitting_eastern45} for (a) the western limb spectrum (at 45$^{\circ}$N-109$^\circ$W).
    }
    \label{fig:app-line_fitting_western45N}
\end{figure}

\begin{figure}
    \centering
    \includegraphics[width=\linewidth]{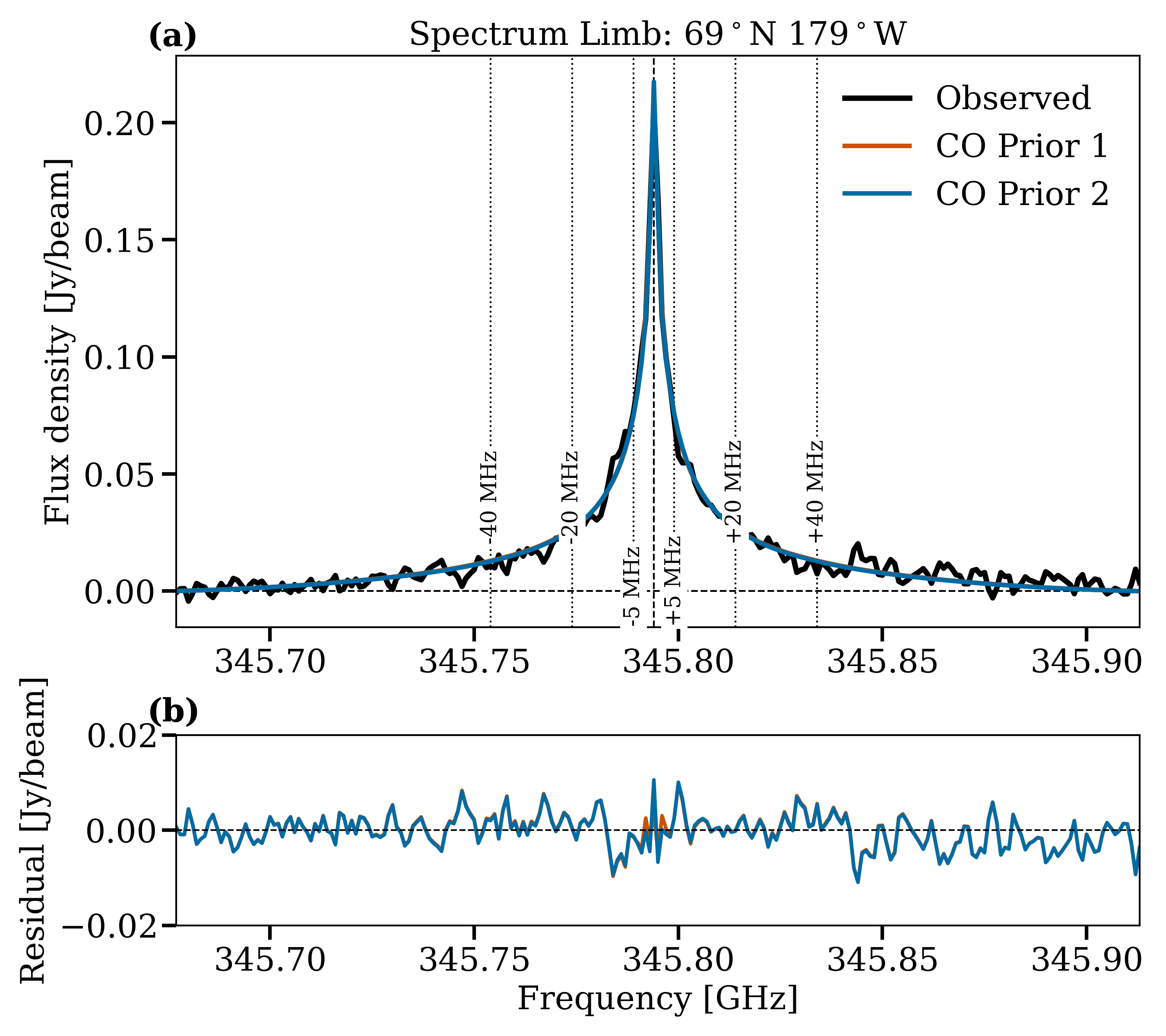}
    \caption{Same as Figure \ref{fig:app-line_fitting_eastern45} for (a) the north polar limb spectrum (at 69$^{\circ}$N-179$^\circ$W).
    }
    \label{fig:app-line_fitting_poleN}
\end{figure}

\FloatBarrier
\onecolumn
\section{Normalized contribution functions}
\label{app:contribution_fonctions}
\begin{figure}[h!]
    \centering
    \includegraphics[width=0.9\linewidth]{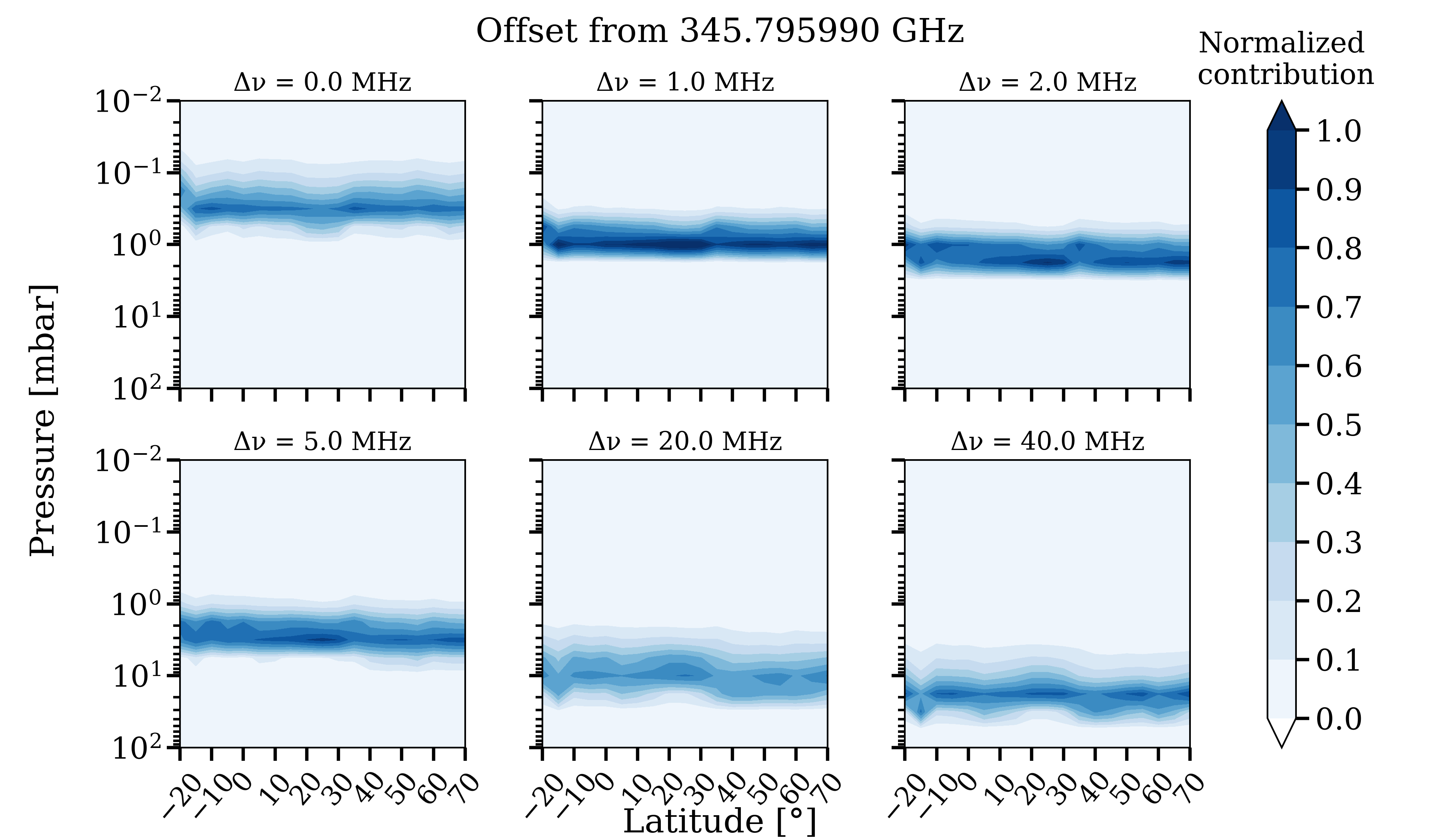}
    \caption{Normalized contribution function for prior 2. 
    }
    \label{fig:app_contribution_function_prior2}
\end{figure}
%\twocolumn

\FloatBarrier
\section{Bin-averaging of meridional variations of CO abundance and error-bars determination}
\label{app:bin-av}
\begin{figure*}
    \centering
    \includegraphics[width=0.8\linewidth]{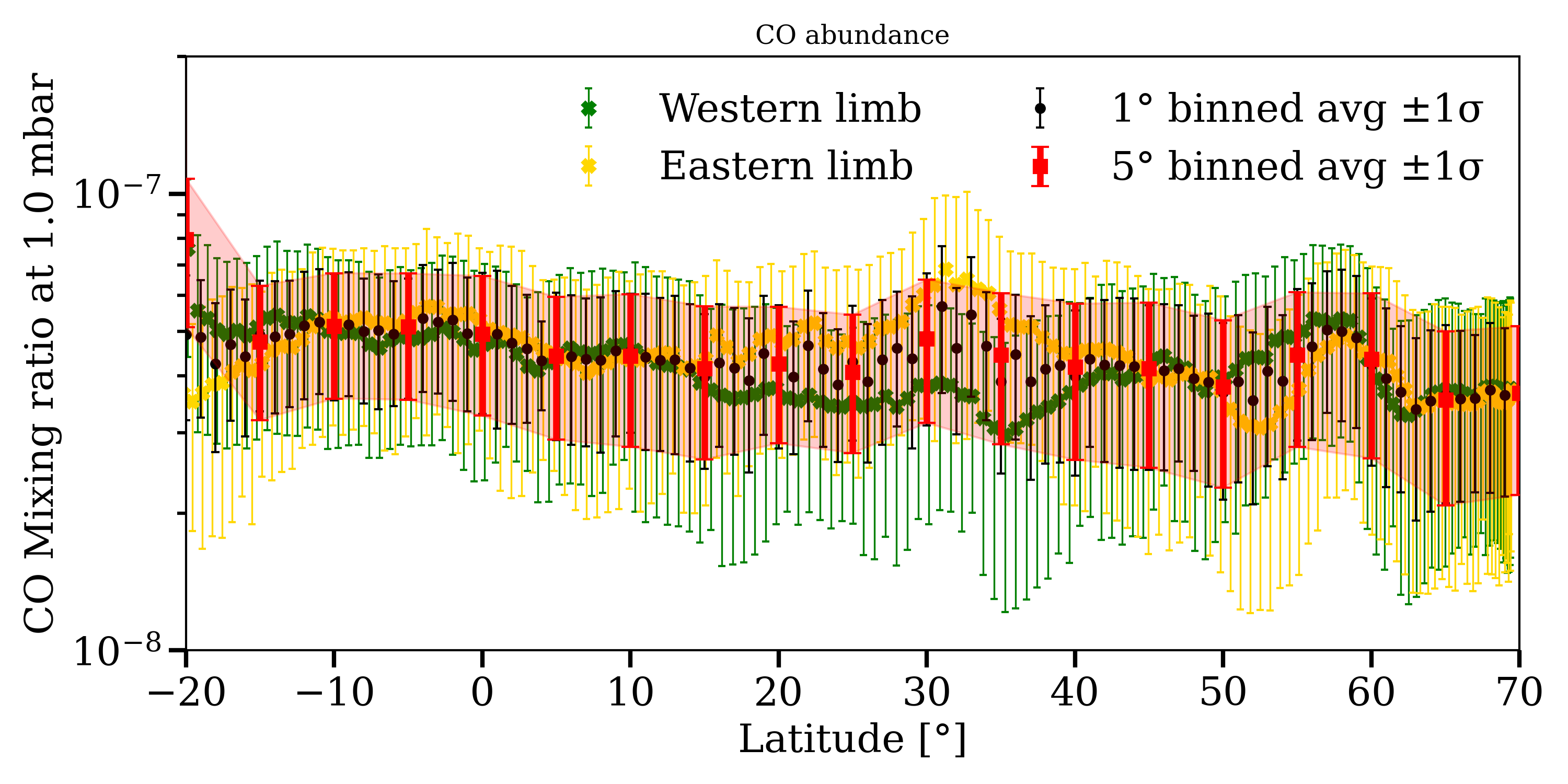}
    \caption{Meridional variations of CO abundance on the west and east limbs at 1 mbar, superimposed with the averaged of the two limbs for each latitude point (in black) and the 5\degr-bin meridional variations (in red) for retrieved CO abundance using the CO Prior 1 profile.}
    \label{fig:app_bin-av_prior1}
\end{figure*}

\begin{figure*}
    \centering
    \includegraphics[width=0.8\linewidth]{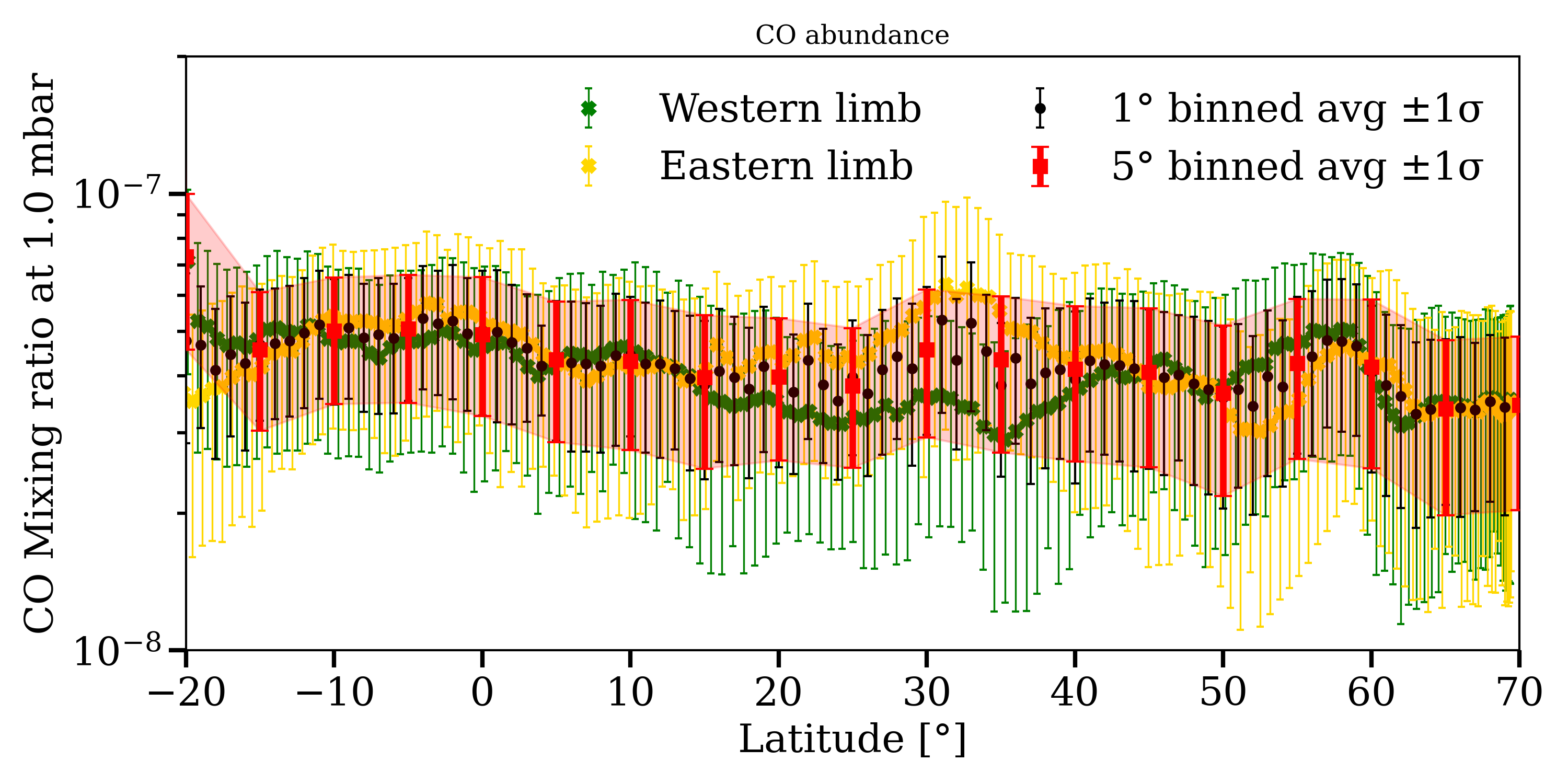}
    \caption{Same as Figure \ref{fig:app_bin-av_prior1} for the CO Prior 2 profile.}
    \label{fig:app_bin-av_prior2}
\end{figure*}

Figures \ref{fig:app_bin-av_prior1} and \ref{fig:app_bin-av_prior2} present the CO retrieved abundance at 1 mbar for both priors on the western and eastern limbs, superimposed with 1$\deg$-bin and 5$\deg$-bin averages of the meridional variation of CO. 
For each latitude point (on the western and eastern limbs), the error-bars are calculated following the method of \cite{Rodg:90}. 
We have, therefore, accounted for the \emph{Null-space} error and the measurement error to determine the uncertainties of our retrieval.
The \emph{Null-space} error corresponds to the error from the smoothing of the vertical profile and the error from the a priori profile, and it is defined as: 
\begin{equation}
    \mathrm{E_1 = (K-I).S.(K-I)^T},
\end{equation}
where $\mathrm{K}$ is the average kernel matrix, $\mathrm{I}$ is the identity matrix, and $\mathrm{S}=e^{-\frac{ln(P_i/P_j)^2}{2L^2}}$ is the smoothing matrix, with $\mathrm{L}$ being the correlation length and $\mathrm{P_i}$ and $\mathrm{P_j}$ the pressures at indexes i and j. 
The measurement error is defined as:
\begin{equation}
    \mathrm{E_2 = U.E.U^T},
\end{equation} where $\mathrm{U=\frac{\alpha S.K^T}{\alpha K.S.K^T + E}}$ is the contribution matrix, $\mathrm{\alpha}$ is the regulation parameter, and $\mathrm{E}$ is the covariance matrix of the measurement noise.

\begin{figure*}
    \centering
    \includegraphics[width=\linewidth]{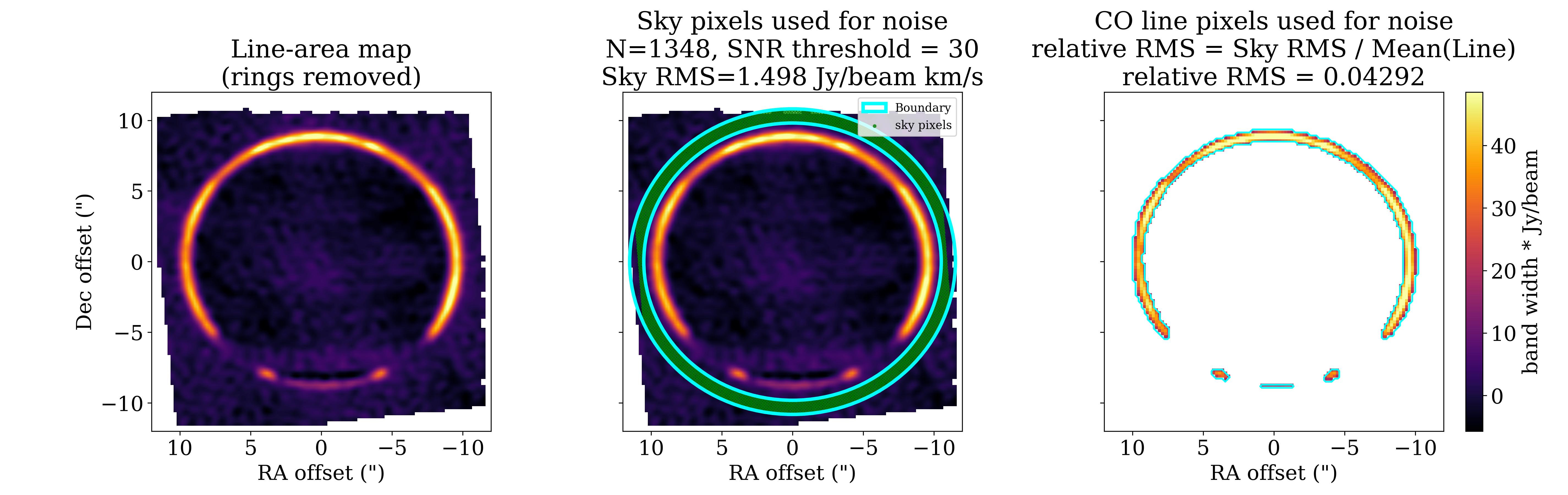}
    \caption{Determination of the relative measurement noise for our ALMA dataset.}
    \label{fig:app_sky_noise}
\end{figure*}

To evaluate the measurement noise, we calculated the relative root mean square (RMS) of the variance of the sky spectra in the image cube presented in Figure \ref{fig:CO_line_intensity_map} (relative to the mean flux of the CO line itself). 
We considered sky spectra to be all spectra in a 1''-width ellipse placed 1.5" away from the limb (as shown in Figure \ref{fig:app_sky_noise}, green shaded area in the middle panel). 
Here, the CO line is isolated thanks to a signal-to-noise ratio threshold of 30: each pixel with a signal 30 times better than the mean flux of the sky pixels is considered a line pixel (right panel in Figure \ref{fig:app_sky_noise}). 
Hence, we were able to evaluate a relative measurement noise of about 0.043, which we applied to determine the error-bars at each latitude point through the covariance matrix of measurement noise $\mathrm{E}$. 
The uncertainty per latitude point is therefore: 
\begin{equation}
    \mathrm{\sigma_{test}=\sqrt{diag(E_1)+diag(E_2)}}
\end{equation}

While two consecutive spectra within a bin of 5$^\circ$-latitude width are not independent (considering the native resolution of the ALMA main array), we averaged all spectra errors in a 5$^\circ$-bin on the eastern and western limbs for each retrieval test, and then we divided the binned mean error by $\sqrt{2}$: 
\begin{equation}
    \mathrm{\sigma_{test,\ average}=\frac{\sigma_{5^{\circ}-bin,\  eastern\ limb}+\sigma_{5^{\circ}-bin,\ western\ limb}}{2 \times \sqrt{2}}}.
\end{equation}
The division by 2 implies that the eastern and western limb measurements are independent, considering that CO abundance is zonally homogeneous. 
Finally, to obtain the mean error-bars for the mean meridional profile of CO abundance, resulting from the average of the results from prior 1 and prior 2 retrievals, we simply calculated 
\begin{equation}
    \mathrm{\sigma_{mean}=\sqrt{\sigma_{test\ 1,\ average}^2+\sigma_{test\ 2,\ average}^2}}.
\end{equation}

\end{appendix}

\end{document}